\documentclass[iop,twocolappendix,trackchanges]{emulateapj}

\usepackage{color}

\usepackage{graphicx}
\usepackage{float}
\usepackage{amsmath}
\usepackage{xcolor}
\usepackage{color}

\usepackage{hyperref}
\usepackage{enumitem}
\setlist{parsep=0pt,listparindent=\parindent}
    {\pagebreak[4]\global\pdfpageattr\expandafter{\the\pdfpageattr/Rotate 90}}%
    {\pagebreak[4]\global\pdfpageattr\expandafter{\the\pdfpageattr/Rotate 0}}%

\newcommand{\JHU}{Department of Physics and Astronomy, The Johns Hopkins University, Baltimore, MD 21218, USA.}
\newcommand{\STScI}{Space Telescope Science Institute, Baltimore, MD 21218, USA.}
\newcommand{\SAAO}{South African Astronomical Observatory, PO Box 9, 7935 Observatory, South Africa}

\newcommand{\AM}{George P. and Cynthia W. Mitchell Institute for Fundamental Physics \& Astronomy, Department of Physics \& Astronomy, Texas A\&M University, College Station, TX, USA}
\newcommand{\UCBerkeley}{Department of Astronomy, University of California, Berkeley, CA 94720-3411, USA}
\newcommand{\Miller}{Miller Senior Fellow, Miller Institute for Basic Research in Science, University of California, Berkeley, CA  94720, USA}

\newcommand{\Duke}{Department of Physics, Duke University, 120 Science Drive, Durham, NC, 27708, USA}
\newcommand{\CapeTown}{Department of Astronomy, University of Cape Town, 7701 Rondebosch, South Africa}

\altaffiltext{1}{\JHU}
\altaffiltext{2}{\STScI}
\altaffiltext{3}{\AM}
\altaffiltext{4}{\SAAO}
\altaffiltext{5}{\CapeTown}
\altaffiltext{6}{\UCBerkeley}
\altaffiltext{7}{\Miller}
\altaffiltext{8}{\Duke}

\newcommand{\apcor}{0.06}
\newcommand{\apcorerr}{0.01}

\newcommand{\lmczpt}{12.21} 
\newcommand{\numstandards}{31}
\newcommand{\nummasterlist}{49,000}

\newcommand{\surfbrightness}{421 }

\newcommand{\slopeuncertainty}{0.01} 

\newcommand{\muone}{31.41}
\newcommand{\mutwo}{31.42}

\newcommand{\pone}{23.15 \pm 0.017} 
\newcommand{\ptwo}{23.18 \pm 0.017} 

\newcommand{\aone}{-6.25 \pm 0.036}
\newcommand{\atwo}{-6.21 \pm 0.036}
\newcommand{\finalsyserr}{0.060}
\newcommand{\absmagsne}{-19.26 \pm 0.120} 
\newcommand{\diffreddening}{0.04}

\newcommand{\finalstaterr}{0.050}

\newcommand{\lmccal}{-6.27} 
\newcommand{\lmcdistmod}{18.477 \pm 0.004 \text{ (stat)} \pm 0.026 \text{ (sys)}}

\newcommand{\HCfullp}{72.7 \pm 4.6}
\newcommand{\HCfullptwo}{72.5 \pm 4.6}
\newcommand{\HCpartp}{74.6 \pm 5.1}
\newcommand{\HCpartptwo}{74.7 \pm 5.1}
\newcommand{\HCLMC}{73.9 \pm 4.3}
\newcommand{\HCLMCtwo}{73.6 \pm 4.3}
\newcommand{\HCCombined}{73.3 \pm 4.0}
\newcommand{\HCCombinedtwo}{73.2 \pm 4.0}

\begin{document}

\title{Hubble Space Telescope Observations of Mira Variables in the Type Ia Supernova Host NGC 1559: An Alternative Candle to Measure the Hubble Constant}
\author{Caroline D. Huang\altaffilmark{1}, Adam G. Riess\altaffilmark{1,2}, Wenlong Yuan\altaffilmark{1}, Lucas M. Macri\altaffilmark{3}, Nadia L. Zakamska\altaffilmark{1}, Stefano Casertano\altaffilmark{2}, Patricia A. Whitelock\altaffilmark{4,5}, Samantha L. Hoffmann\altaffilmark{2}, Alexei V. Filippenko\altaffilmark{6,7}, and Daniel Scolnic\altaffilmark{8}}

\begin{abstract} 

We present year-long, near-infrared \emph{Hubble Space Telescope} WFC3 observations used to search for Mira variables in NGC 1559, the host galaxy of the Type Ia supernova (SN Ia) 2005df. This is the first dedicated search for Miras, highly-evolved low-mass stars, in a SN~Ia host and subsequently the first calibration of the SN Ia luminosity using Miras in a role historically played by Cepheids. We identify a sample of 115 O-rich Miras with $P < 400$ day based on their light curve properties.  We find that the scatter in the Mira Period--Luminosity Relation (PLR) is comparable to Cepheid PLRs seen in SN~Ia supernova host galaxies.  Using a sample of O-rich Miras discovered in NGC 4258 with \emph{HST} \emph{F160W} and its maser distance, we measure a distance modulus for NGC 1559 of $\mu_{1559} = \muone \pm \finalstaterr $ (statistical) $\pm \finalsyserr $ (systematic) mag. Based on the light curve of the normal, well-observed, low-reddening SN 2005df, we obtain a measurement of the fiducial SN Ia absolute magnitude of $M_B^0= -19.27 \pm 0.13 $ mag.  
With the Hubble diagram of SNe~Ia we find $H_0 = \HCfullp$ km\,s$^{-1}$\,Mpc$^{-1}$. 
Combining the calibration from the NGC 4258 megamaser and the Large Magellanic Cloud detached eclipsing binaries gives a best value of $H_0 = \HCCombined$ km\,s$^{-1}$\,Mpc$^{-1}$. This result is within 1$\sigma$ of the Hubble constant derived using Cepheids and multiple calibrating SNe~Ia. This is the first of four expected calibrations of the SN Ia luminosity from Miras which should reduce the error in $H_0$ via Miras to $\sim$ 3\%. In light of the present Hubble tension and \emph{JWST}, Miras have utility in the extragalactic distance scale to check Cepheid distances or calibrate nearby SNe in early-type host galaxies that would be unlikely targets for Cepheid searches. 

\end{abstract}

\section{Introduction}
\label{sec:intro}
One of the most intriguing issues in observational cosmology is the discrepancy between the Hubble Constant ($H_0$) measured in the present-day Universe using the Cepheid-supernovae distance ladder \citep[][hereafter, R16]{Riess16} and gravitational lensing time delays \citep{Bonvin17, Wong19} and the one inferred from the early Universe using the cosmic microwave background (CMB) data assuming a $\Lambda$CDM cosmology \citep{Planck18VI}. Combining ind ependent \emph{SH0ES} \citep[][hereafter, R19]{Riess19} and H0LiCOW measurements (for a recent review, see \citet{Verde19} and sources within), the difference is at the $5.3\sigma$ confidence level. Despite increasing attention, it appears difficult to explain the discrepancy as being caused by a single error in any specific measurement \citep{Addison18, Aylor19}. The persistence of the tension has led to the possibility of additions to $\Lambda$CDM as an explanation (e.g., \cite{Poulin19}), but the standard of proof for new physics is high, and alternative routes to the present measurements should be explored. 

Independent, intermediate range distance indicators such as highly-evolved Mira variable stars and the Tip of the Red Giant Branch (TRGB) \citep{Jang17, Freedman19, Yuan19} can help provide a check of local Cepheid distances. They can also increase the number of local Type Ia supernovae (SNe~Ia) used as calibrators whose sample size limits the precision of $H_0$. Though evolved, Cepheids have massive progenitors and are thus relatively young stars, so \emph{HST} searches for such variables in local SNe hosts target nearby, globally star-forming, late-type galaxies with modest inclination. Miras have absolute magnitudes comparable to those of Cepheids in the near-infrared (NIR) and can be seen out to the same volume, allowing them to fulfill a role in the distance scale similar to that of Cepheids, as calibrators of SNe~Ia. Unlike Cepheids, short-period ($P < 400$ days) Miras have low-mass progenitors and are present in all galaxy morphologies \citep{Boyer17, Whitelock14, Rejkuba04}. Miras can also be found in the halos of galaxies, removing the need for low-inclination hosts. The ubiquity of Miras makes them a potentially attractive tool to calibrate SNe~Ia in all galaxy types.

Though Miras were the first pulsating variable stars discovered (over 400\,yr ago), they remain relatively underused as a distance indicator. They are asymptotic giant branch (AGB) stars that pulsate in the fundamental mode with periods ranging from $\sim 100$ days on the short end to greater than $2000$ days for the longest-period objects. They have amplitude greater than 2.5 mag in the \emph{V} band, the largest of any regular pulsator \citep{Soszynski05}. Miras were suggested to have Period--Luminosity Relations (PLRs) by \citet{Gerasimovic28}, but the scatter in visual magnitudes at a given period was large ($\sim 0.5$ mag), as summarized by \citet{Feast89}.  Later studies \citep{GlassLE81, Robertson81} using observations in the NIR found a much tighter PLR, suggesting that they could be promising distance indicators at longer wavelengths.  They are typically divided into two subclasses, oxygen-rich (O-rich) and carbon-rich (C-rich) \citep[][see \citet{Ita11} for comparisons of their PLRs]{Kleinmann79}. The O-rich Miras follow a well-defined PLR at IR wavelengths, while the C-rich Miras fall fainter than the PLR in the optical and NIR, likely owing to circumstellar reddening. However, corrected for reddening in the K-band (2.2~$\mu$m) C-rich Miras and O-rich Miras start to follow a similar PLR \citep{Whitelock03}. 

Host-galaxy metallicity affects Miras in several ways. Galaxies with higher metallicity exhibit larger ratios of O-to-C-rich Miras \citep{Boyer17, Hamren15, Battinelli05}. In addition, O-rich Miras with more massive progenitors ($\sim 3\, M_\odot$; $P \gtrsim 400$ days) can experience hot-bottom-burning (HBB) \citep{Whitelock03, Marigo17}. HBB stars are brighter at their period range than would be predicted from a linear PLR fit to shorter-period objects. Though there is no consensus on the details of HBB, the onset of HBB is known to have a dependency on both mass (and thus period, since more-massive Miras have longer periods) and metallicity. Therefore, for distance measurements, selecting Miras with $P< 400$ days removes the known metallicity dependences of the Mira PLR. 

Miras are not as commonly used as intermediate range (5-50 Mpc) distance indicators as Cepheids or TRGB because obtaining the necessary observations has been more difficult. Years-long baseline observations are necessary in order to obtain accurate measurements of their periods. Furthermore, they do not follow tight PLRs in the optical --- only in the IR --- and these observations are more difficult to obtain from the ground. Mira light curves can be observed in the optical to obtain accurate period measurements, but still need IR follow-up observations in order to produce useful PLRs. 

Until recently, calibrating their luminosities has also been challenging.
Previous studies of Miras have used the Large Magellanic Cloud \citep[LMC;][]{Soszynski13, Ita04, Feast89} and M33 \citep{Yuan17a} to calibrate the luminosity of these stars. \citet{Whitelock08} showed using Hipparcos parallaxes that the PLR for Galactic Miras is consistent with that in the LMC. Individual parallaxes in the Milky Way are more difficult to measure than for most stars because the angular sizes of Miras are larger than their parallaxes (their physical sizes are larger than 1\,au).  In addition,  they typically have time-variable asymmetries \citep{Paladini17} speculated to result from giant convection cells \citep{Lattanzi97} or caused by nonradial pulsations. Even with \emph{Gaia}, obtaining Mira parallaxes may only be possible with later data releases that can properly account for movement of the photocenter and the resolved profiles of these sources \citep{Whitelock18}. 

In our previous paper \citep[][hereafter, H18]{Huang18}, we addressed the issue of calibration by searching for Miras in the water megamaser host galaxy NGC 4258. In addition to providing us with the farthest Miras to have measured luminosities and periods, the host galaxy's water megamaser allowed for a geometric distance with 2.6\% precision (R16; \citealt{Humphreys13}). This enabled us to obtain an absolute calibration for the Mira distance scale on the \emph{HST} system of filters. These results were in good agreement with the LMC Mira calibration of \citet{Yuan17b}. The maser distance of NGC 4258 has recently improved to 1.5\% precision, adding more precision to this calibration \citep{Reid19}.

In this paper, we present the first observations of Miras in a SN~Ia host, yielding the first Mira-based calibration of the luminosity of SNe~Ia. It is the first of four local SN Ia hosts in which we are searching for Miras with \emph{HST}. In Section \ref{sec:data} we describe the observations, data reduction, and photometry. The selection criteria for Miras are discussed in \ref{sec:selection}. In Section \ref{sec:sys} we characterize the systematic uncertainties. We present and discuss the main result in Section \ref{sec:results} and conclude in Section \ref{sec:conclusions}. Throughout the paper we use ``amplitude" to refer to the total (peak-to-trough) variation of a variable star light curve over the course of one cycle. 

\begin{deluxetable*}{lcccc}
\tabletypesize{\scriptsize}
\tablecaption{{\it HST}-GO 15145 Observations Used in this Work}
\tablewidth{0pt}
\tablehead{\colhead{Epoch} & \colhead{Orientation (deg)} & \colhead{MJD} & \colhead{UT Date}}
\startdata

01 & 160.6 & 58005.2 & 2017-09-08 \\
02 & 160.6 & 58015.8 & 2017-09-19 \\
03 & $-165.4$ & 58037.6 & 2017-10-11 \\
04 & $-165.4$ & 58049.3 & 2017-10-22 \\
05 & $-124.4$ & 58084.9 & 2017-11-27 \\
06 & $-78.6$ & 58142.2 & 2018-01-23 \\
07 & $-18.1$ & 58199.3 & 2018-03-21 \\
08 & 36.6 & 58256.2 & 2018-05-17 \\
09 & 90.3 & 58314.6 & 2018-07-15 \\
10 & 160.6 & 58372.8 & 2018-09-11 
\enddata
\tablecomments{The WFC3/IR camera and \emph{F160W} filter were used for each observation. The exposure time of each epoch was 1006\,s. The orientiation is the angle east of north of the $y$-axis of each image.} 
\label{tab:obs}
\end{deluxetable*}

\section{Observations, Data Reduction, and Photometry}
\label{sec:data}

\subsection{Observations and Data Reduction}
\label{sec:obs}

This study uses ten epochs of \emph{HST} WFC3/IR \emph{F160W} data (GO-15145; PI Riess), with an exposure time of 1006\,s for each epoch. The observations are also designed to measure Cepheid variables found using epochs of optical imaging (Miras are too red to appear in the optical images).  The epochs have an approximately monthly cadence between 2017 September 8 and 2018 September 11, with a field of view centered at $\alpha = 04^{\text{h}}17^\text{m}37^\text{s}$ and $\delta = -62^\circ47'00''$ (J2000). Table \ref{tab:obs} contains the observational information for each epoch. 

We use pipeline-calibrated images retrieved from the \emph{HST} MAST Archive. We generate pixel-resampled and stacked images for each epoch using {\tt Drizzlepac 2.2.2} \citep{Gonzaga12}. Each epoch contains two subpixel dither positions and has been resampled to a scale of $0.12''$ pix$^{-1}$ (WFC3 has a native scale of $0.128''$ pix$^{-1}$) and the orientation of the images rotates by slightly over a full rotation as a consequence of \emph{HST} orient constraints. We choose the first epoch as the reference image and align all of the subsequent images with it using {\tt DrizzlePac}.

\subsection{Photometry and Calibration}
\label{sec:phot}

\setlength{\tabcolsep}{1em}
\begin{deluxetable*}{lccrrcc}
\tabletypesize{\scriptsize}
\tablecaption{Secondary Standards}
\tablewidth{0pt}
\tablehead{\colhead{ID} & \colhead{$\alpha$} & \colhead{$\delta$} & \colhead{$X$} & \colhead{$Y$} & \colhead{$F160W$} & \colhead{Uncertainty} \\
\colhead{} & \colhead{(J2000)} & \colhead{(J2000)} & \colhead{(pixels)} & \colhead{(pixels)} & \colhead{(mag)} & \colhead{(mag)} 
}

\startdata

     7599 &   04$^{\rm h}17^{\rm m}$40.019$^{\rm s}$ & -62$^\circ 46'17.51''$ & 843.0 & 203.9 & 18.939 & 0.081 \\ 
    10534 &   04 17 34.146   &     -62 46 41.15 &  461.3 &  280.8 &   19.039 &    0.078 \\ 
     4002 &   04 17 31.105   &     -62 46 26.79 &  335.7 &  111.1 &   19.195 &    0.070 \\ 
    25011 &   04 17 37.447   &     -62 47 23.13 &  525.8 &  673.1 &   19.238 &    0.093 \\ 
    10962 &   04 17 39.975   &     -62 46 28.86 &  809.8 &  292.5 &   19.355 &    0.104 \\ 
     8924 &   04 17 37.077   &     -62 46 28.90 &  653.0 &  238.8 &   19.381 &    0.060 \\ 
    25548 &   04 17 32.556   &     -62 47 36.57 &  225.1 &  687.9 &   19.433 &    0.066 \\ 
    16153 &   04 17 39.255   &     -62 46 48.42 &  717.8 &  433.2 &   19.466 &    0.107 \\ 
     4308 &   04 17 29.008   &     -62 46 32.80 &  206.1 &  119.5 &   19.684 &    0.070 \\ 
    31503 &   04 17 41.974   &     -62 47 34.16 &  740.5 &  844.3 &   19.789 &    0.083 \\ 
    25269 &   04 17 37.822   &     -62 47 23.10 &  546.2 &  679.8 &   19.823 &    0.050 \\ 
    11364 &   04 17 38.388   &     -62 46 34.00 &  710.0 &  303.5 &   19.895 &    0.052 \\ 
     5369 &   04 17 40.051   &     -62 46 10.18 &  864.6 &  146.8 &   19.986 &    0.074 \\ 
    24951 &   04 17 44.680   &     -62 47 05.78 &  963.8 &  671.1 &   20.022 &    0.064 \\ 
     7243 &   04 17 38.170   &     -62 46 20.78 &  734.1 &  195.2 &   20.168 &    0.075 \\ 
    29515 &   04 17 41.245   &     -62 47 29.25 &  714.4 &  791.0 &   20.193 &    0.111 \\ 
    26819 &   04 17 32.001   &     -62 47 42.17 &  179.9 &  721.8 &   20.195 &    0.064 \\ 
    25967 &   04 17 42.801   &     -62 47 13.83 &  840.4 &  699.5 &   20.236 &    0.166 \\ 
    31273 &   04 17 35.870   &     -62 47 47.77 &  373.8 &  837.8 &   20.236 &    0.077 \\ 
     1510 &   04 17 34.915   &     -62 46 09.25 &  589.4 &   43.7 &   20.242 &    0.057 \\ 
    35832 &   04 17 43.821   &     -62 47 43.95 &  813.7 &  955.8 &   20.265 &    0.094 \\ 
    23326 &   04 17 42.191   &     -62 47 06.19 &  828.2 &  627.9 &   20.282 &    0.055 \\ 
    29323 &   04 17 32.986   &     -62 47 48.13 &  216.9 &  787.1 &   20.325 &    0.059 \\ 
    30041 &   04 17 34.260   &     -62 47 47.51 &  287.5 &  805.9 &   20.659 &    0.109 \\ 
    27599 &   04 17 37.056   &     -62 47 32.79 &  478.5 &  741.9 &   20.690 &    0.082 \\ 
    27351 &   04 17 34.181   &     -62 47 38.82 &  306.8 &  735.9 &   20.702 &    0.067 \\ 
    28149 &   04 17 33.826   &     -62 47 42.35 &  278.0 &  757.1 &   20.778 &    0.078 \\ 
    35037 &   04 17 39.297   &     -62 47 51.99 &  547.5 &  934.9 &   20.803 &    0.063 \\ 
     9301 &   04 17 29.739   &     -62 46 47.50 &  205.8 &  248.9 &   20.835 &    0.075 \\ 
     5013 &   04 17 40.445   &     -62 46 08.02 &  891.7 &  137.1 &   20.932 &    0.067 \\ 
     6365 &   04 17 38.468   &     -62 46 17.28 &  759.7 &  173.2 &   21.021 &    0.059 
\enddata
\tablecomments{All of the secondary sources used to calibrate the \emph{F160W} light curves. ID numbers are photometry IDs. The $X$ and $Y$ positions are relative to the first epoch of the \emph{F160W} image, and the uncertainties are photometric errors as estimated by {\tt DAOPHOT}. The magnitudes are calculated as explained in Eq. \ref{eq:mag}.} 
\label{tab:secondarystandards}
\end{deluxetable*}

We begin by stacking all of the \emph{F160W} observations to make a deeper ``master" image. This image is used to create a master source list for the photometry and also to derive crowding used in period-fitting. Given the crowded nature of our fields, we use the crowded-field photometry packages {\tt DAOPHOT/ALLSTAR} \citep{Stetson87} and {\tt ALLFRAME} \citep{Stetson94} for the data reduction and source extraction. Our procedure closely follows the steps used to search for Miras in the water megamaser host NGC 4258 (H18).

The {\tt DAOPHOT} routine {\tt FIND} is used to detect sources with a $>3\sigma$ significance. The significance depends on sky background, readout noise, and the number of images combined. Then the {\tt DAOPHOT} routine {\tt PHOT} is used to perform aperture photometry. We input the star list produced from the aperture photometry into {\tt ALLSTAR} for point-spread-function (PSF) photometry, using a 2.5 pixel full-width at half-maximum intensity (FWHM), standard input into {\tt ALLSTAR}. We then repeat these steps on the star-subtracted image generated by {\tt ALLSTAR} (with all of the previously-discovered sources removed) to produce a second source list. The two source lists are then concatenated and used as input into another round of {\tt ALLSTAR} (applied to the image without subtracted sources) to create a final master source list of $\sim$ \nummasterlist \ objects. 

This master source list is input into {\tt ALLFRAME}. We use the master source list to derive a consistent source list for PSF photometry of each \emph{F160W} epoch. {\tt ALLFRAME} is similar to {\tt ALLSTAR}, except that it is capable of simultaneously fitting the profiles of all sources across the full baseline of epochs for a single field. This allows it to maintain a constant source list through multiple epochs and improves the photometry for stars closer to the detection limit. {\tt ALLFRAME} produces time-series PSF photometry as output. 

Secondary standards are used to correct for variations in the photometry of constant sources (detector and image quality) between different epochs. We search for secondary standards in the star lists by choosing bright objects that were observed in all ten \emph{F160W} epochs. Their stellar profiles and surroundings were visually inspected to choose secondary standards that are relatively isolated compared to other stars in the image, removing any that showed variability or had large photometric errors. This leaves us with a total of \numstandards \, sources, summarized in Table \ref{tab:secondarystandards}. We calculate the celestial coordinates for all of these sources using the astrometric solutions in the FITS headers. Photometry is conducted on the resampled images. The mean residuals for these stars across all epochs of \emph{F160W} imaging exhibited a dispersion of 0.01 mag. We use the standard photometry routine {\tt TRIALP} (kindly provided by Peter Stetson) to derive the epoch-to-epoch changes in zeropoint using our standard stars and this is used to correct the {\tt ALLFRAME} photometry for epoch-dependent variations.

To put our observations on the same zeropoint as H18, which we adopt for an absolute calibration of distance, we use the same \emph{HST} WFC3 IR photometric zeropoint for a $0.4''$-radius aperture. In Vega magnitudes, the \emph{F160W} zeropoint is 24.5037 mag.  The use of a consistent zeropoint in a fixed aperture where we normalize the flux of the PSF compensates for small differences on the observing pattern, PSF, and resampling scale between NGC 4258 and NGC 1559.  In NGC 4258, the observations of each epoch consisted of four subpixel dithers. This allowed us to drizzle images for each epoch to a finer scale in NGC 4258 ($0.08''$ pix$^{-1}$) than in NGC 1559 ($0.12''$ pix$^{-1}$), where we only have two subpixel dithers per epoch. 

To define aperture corrections to the PSF, we select the brightest and most isolated stars in the field that are not saturated. These are often foreground stars, unlike secondary standards, which are chosen to be closer in expected magnitude to our variables. These stars are used to determine the correction from the PSF magnitude to the $0.4''$ aperture magnitude only, not as photometric standards.

We subtract all other stars in the master source image from the image using {\tt DAOPHOT/SUBSTAR} to create a ``clean" image.  Aperture photometry is then performed on the bright stars in the clean image using 10 apertures ranging from 1 to 5 pixels to create a growth curve. We inspect the growth curve of each of the stars used in the aperture corrections and reject any stars that do not have smooth, monotonically increasing growth curves. The aperture correction is defined as
\begin{align}
\Delta m_{\text{ac}} = m_{\text{ap}} - m_{\text{PSF}},
\end{align}
where $m_{\text{ap}}$ is the aperture magnitude for an aperture with a radius of 3.3 pixels, equivalent to the $0.4''$ aperture zeropoint, and $m_{\text{PSF}}$ is the PSF magnitude. We find an aperture correction of $\apcor \pm \apcorerr$ magnitudes, which we add to $m_{\text{PSF}}$ to correct to the \emph{HST} zeropoint for a $0.4''$ aperture. PSF photometry is used throughout our analysis, and the aperture correction is included at the end to put our old and new observations on the same \emph{HST} zeropoint. The aperture correction may change slightly between the observations of NGC 4258 and NGC 1559 owing to ``breathing" and small focus changes. Because we do not have standard stars in common between our two fields, we cannot correct for this effect, which is estimated to be $\sim 0.01$\,mag, so we include this in the uncertainty. Thus, including the aperture correction, we convert from instrumental magnitudes to calibrated magnitudes using
\begin{equation}
\label{eq:mag}
m_f = m_i - 2.5\log(t) - a_{\text{DAO}} + a_{\text{HST}} + \Delta m_{\text{ac}} + \Delta m_b,
\end{equation}
where $m_f$ is the final calibrated magnitude, $m_i$ is the instrumental magnitude output from {\tt DAOPHOT} photometry, $t$ is the exposure time, $a_{\text{DAO}}$ is the {\tt DAOPHOT} photometric zeropoint (25.0 mag), $a_{\text{HST}}$ is the \emph{HST} zeropoint for a $0.4''$ aperture (24.504 mag), $\Delta m_{\text{ac}}$ is the aperture correction (0.06 mag), and $\Delta m_b$ is the crowding bias correction (mean value of $0.13$ mag) discussed in \S \ref{sec:surfbrightness}. 

\section{Mira Selection Criteria}
\label{sec:selection}

Light-curve templates are typically used to identify and classify variable stars. Classical Cepheids, RR Lyrae stars, and Type II Cepheids all have distinctive light-curve shapes, and templates for these objects exist in both the optical and the NIR \citep{Jones96, Yoachim09, Sesar10, Inno15, Bhardwaj17}. However, unlike most other variable stars that have been used as distance indicators, Miras do not have a homogeneous light-curve shape. In addition, both the shape and magnitude at a given phase of Mira light curves are known to vary between cycles, though these variations are smaller at NIR wavelengths \citep{Whitelock94, Olivier01, Yuan17b}.  Therefore, a sine function has been used to fit NIR Mira light curves in previous studies \citep{Matsunaga09}.  Here we also use a single sine function to fit the periodicity, and the amplitude of the sine to classify the variables. While the amplitude may change in subsequent cycles, the period remains more consistent, so a sine fit still allows us to capture the periodicity. 

We use a number of steps to identify Miras and separate candidates from other types of Long-Period Variables (LPVs) and non-variable stars, including the Stetson variability index $L$, period, amplitude, and an approximate $F$-statistic. All of the candidate variables that pass these criteria are also visually inspected to remove any potential noisy stars or non-variable contaminants. In addition, use information about the local environment of the candidate Miras --- surface brightness and crowding bias corrections --- to determine if they are reliable sources not dominated by background flux. Finally, we use simulations (described in Section \ref{sec:sims}) to test the efficacy of our selection criteria. 

The final sample after applying all of our selection criteria outlined in \S \ref{sec:variability}, \S \ref{sec:hamp}, \S \ref{sec:percut}, and \S \ref{sec:surfbrightness} is shown in Table \ref{tab:miratable}.

\subsection{Variability and Period Determination}
\label{sec:variability}

\setlength{\tabcolsep}{.7em}
\begin{deluxetable*}{lcccrrcccc}
\tabletypesize{\scriptsize}
\tablecaption{Final Sample of Miras in NGC 1559}
\tablewidth{0pt}
\tablehead{\colhead{ID} & \colhead{Period} & \colhead{$\alpha$} & \colhead{$\delta$} & \colhead{$X$} & \colhead{$Y$} & \colhead{Mag} & \colhead{$\sigma$} & \colhead{Amp} & \colhead{$\Delta m_b$}\\
\colhead{} & \colhead{days} & \colhead{J2000} & \colhead{J2000} & \colhead{pix} & \colhead{pix} & \colhead{\emph{F160W}} & \colhead{\emph{F160W}}  & \colhead{$\Delta$ mag} & \colhead{mag}
}
\startdata
  300258 &  374.586 &   04$^{\rm h}17^{\rm m}36.487^{\rm s}$  &     -62$^\circ 46' 00.66''$ &  697.660 &    5.385 &   24.250 &    0.153 &    0.674  & 0.002 \\
      816 &  344.937 &   04 17 37.056   &     -62 46 01.59 &  725.941 &   23.264 &   24.302 &    0.163 &    0.610 &    0.072 \\
   300938 &  272.999 &   04 17 34.200   &     -62 46 08.26 &  553.360 &   22.640 &   24.749 &    0.176 &    0.605 &    0.181 \\
      934 &  339.024 &   04 17 32.541   &     -62 46 12.72 &  451.567 &   26.954 &   25.159 &    0.184 &    0.637 &    0.013 \\ 
     1372 &  260.733 &   04 17 37.048   &     -62 46 03.75 &  719.626 &   40.124 &   24.328 &    0.270 &    0.744 &    0.032 \\
     1423 &  240.849 &   04 17 28.711   &     -62 46 23.70 &  214.651 &   42.213 &   24.118 &    0.178 &    0.654 &    0.011 \\
   301623 &  295.503 &   04 17 37.568   &     -62 46 03.07 &  749.607 &   44.512 &   25.016 &    0.253 &    0.580 &    0.080 \\
     1964 &  263.733 &   04 17 34.958   &     -62 46 10.78 &  587.545 &   56.643 &   24.359 &    0.473 &    0.475 &    0.073 \\
     1606 &  310.348 &   04 17 30.305   &     -62 46 20.60 &  309.246 &   47.456 &   24.363 &    0.145 &    0.465 &    0.070 \\
     2227 &  347.314 &   04 17 33.852   &     -62 46 14.37 &  517.991 &   64.354 &   24.607 &    0.183 &    0.649 &    0.044 \\
     2640 &  381.497 &   04 17 38.749   &     -62 46 04.18 &  810.458 &   75.226 &   22.735 &    0.169 &    0.581 &    0.033 \\
   303038 &  337.114 &   04 17 32.496   &     -62 46 19.88 &  429.738 &   82.513 &   25.765 &    0.160 &    0.767 &    0.068 \\
   303532 &  319.410 &   04 17 31.855   &     -62 46 22.97 &  386.645 &   94.986 &   24.882 &    0.216 &    0.686 &    0.095 \\
     3973 &  272.939 &   04 17 36.615   &     -62 46 13.59 &  669.508 &  109.656 &   24.565 &    0.372 &    0.650 &    0.225 \\
     3626 &  251.081 &   04 17 30.908   &     -62 46 25.97 &  327.299 &  101.019 &   25.140 &    0.261 &    0.675 &    0.095 \\
     4015 &  326.046 &   04 17 38.073   &     -62 46 10.34 &  757.175 &  111.162 &   24.673 &    0.363 &    0.461 &    0.155 \\
     4099 &  267.813 &   04 17 38.127   &     -62 46 10.41 &  759.922 &  112.767 &   24.921 &    0.484 &    0.449 &    0.144 \\
   304685 &  254.121 &   04 17 37.465   &     -62 46 13.60 &  715.432 &  125.538 &   23.670 &    0.485 &    0.544 &   -0.011 \\
   304909 &  328.070 &   04 17 28.879   &     -62 46 34.53 &  194.400 &  130.754 &   24.537 &    0.216 &    0.500 &    0.017 \\
     5985 &  257.784 &   04 17 28.723   &     -62 46 39.15 &  173.463 &  164.217 &   24.462 &    0.215 &    0.589 &    0.044 

\enddata
\setlength{\tabcolsep}{0.7em}
\tablecomments{A partial list of Miras is shown here for information regarding form and content. The final magnitudes are consistent with Eq. \ref{eq:mag}. In order to fit the PLR we also applied an extinction correction of $-0.01$ mag and a C-rich contamination correction of $-0.057$ mag. To obtain the magnitudes in Figure \ref{fig:plrlinear}, add $-0.067$ mag to the magnitude column of this table.} 
\label{tab:miratable}
\end{deluxetable*}

\begin{figure}
\epsscale{1.2}
  \plotone{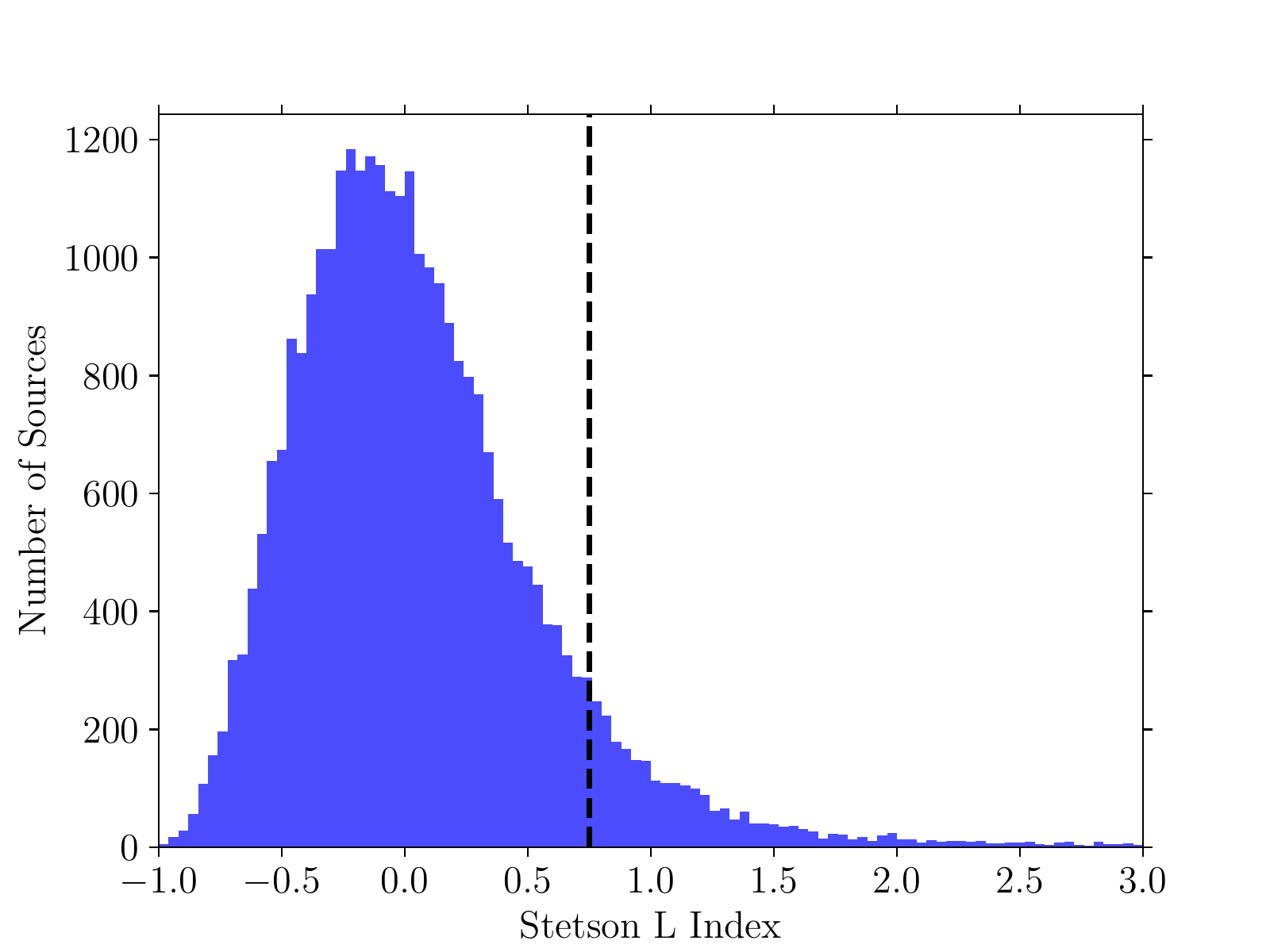}
  \caption{ The distribution of the Stetson $L$ index for all of the objects detected in the master image in ten epochs of observations. We began the variability search by examining objects with $L \geq 0.75$, indicated in the plot by the black dashed line.  There were $\sim \nummasterlist$ objects in the master list and $\sim 25,000$ with 10 epochs of observations.  
   }
  \label{fig:stetsonhist}
\end{figure}

The first step in our Mira selection process is calculating the Welch-Stetson variability index \emph{L} of all of the objects in our field \citep{Stetson96}. This index is only used to weed out as many nonvariable stars as possible before beginning the later parts of the search, as many objects with high variability indexes may still be non-Miras, short-period variables, or simply nonvariable stars with noisy measurements. We use a cutoff for the minimum variability index of $L \geq 0.75$, consistent with the criterion H18 used for NGC 4258. With simulations, we find that this variability cut removes $\sim 90\%$ of the nonvariable stars. Figure \ref{fig:stetsonhist} shows a distribution of the $L$ index for all of the stars we detect in our master image. Approximately 3000 variable-star candidates remain at this stage. 

Next, we calculate the most likely period (assuming the variability is periodic) of the remaining candidates. For each, we perform a grid search on periods ranging from 100 to 1000 days, the relevant period range for Miras. At each period we fit the light curve in magnitude space with a sine function with the amplitude, mean magnitude, and phase as free parameters using a $\chi^2$ minimization, as done by H18. In addition, the mean magnitude from the sine fit is used as the mean magnitude of the Mira for fitting the PLR in both this paper and H18, giving us a consistent definition of mean magnitude.  The period with the lowest $\chi^2$ fit to a sine is then chosen as the best-estimate period of the Mira. As our observation baseline spanned $\sim 370$ days, and the shortest-period Miras we can effectively use in our PLR relations have $P = 240$ days (see \S 3.3), we would not expect to see aliasing of periods within in our target range in our period estimations. Our observation timing is such that each Mira in our period range is sampled at least five phases per cycle. 

Aliasing of shorter period variables to the period range of Miras that we seek is possible. However, this should be rare since we an amplitude cut to our sample, removing objects that do not have large amplitudes in \emph{F160W}. The majority of variable stars with shorter periods (such as Cepheids) that could be aliased with our monthly observing-cadence also have amplitudes $<0.4$ mag in the NIR and would not pass these amplitude cuts (see Figure 2 of H18 for a comparison of $H$-band amplitudes of variable stars in the LMC). \cite{Laney93} found that some Cepheids with periods between 20 and 40 days may reach an amplitude of 0.4 mag in the $H$-band. However, these Cepheids would have massive progenitors and are expected to be less common in the lower surface-brightness regions of the galaxy we exclude most objects from the star-forming regions of the galaxy, in our surface brightness cuts, explained in \S \ref{sec:surfbrightness}. 

Aliasing of longer-period variables ($P > 400$ days) into our sample period range is also possible---particularly if we discover a Mira with a long secondary period and a bump. This bump can resemble the end of one cycle if we do not have a long enough baseline to observe the full light curve for one of the Mira's pulsational cycle. However, these stars are typically C-rich, heavily reddened, and more massive ($\gtrapprox 3M_{\odot}$), thus making them relatively rare. A cursory examination of the OGLE data yields approximately $\sim 60$ stars out of a sample of $> 1500$ ($\sim 4\%$ of the sample) that have these qualities for the entire LMC sample. Most of them clustered in the bar of the LMC, further suggesting that they are likely to have massive progenitors. In addition, lightcurves---especially those of heavily-reddened stars---are more regular in the NIR than they are in optical wavelengths, so we also expect this effect to be smaller than in the optical OGLE data. 

Finally, we refer to H18, where we were able to combine optical observations of Mira variables with a rapid cadence of about one visit every few days (used to search for Cepheids) and a baseline of $\sim 40$ days with the roughly monthly visits and year-long baseline used to search for Miras. In our gold sample of Miras, which contained objects with both \emph{F814W} lightcurves and \emph{F160W} lightcurves, the zeropoint was within $0.02$ mag of that of the bronze sample Miras that had only \emph{F160W} lightcurves to classify them, as is the case in NGC 1559. In addition, all of the optical light curves we were able to recover for our sources indicated long-period objects consistent with the periods we determined in our analysis. Additionally, phase near the trough of the light curve at the time of the optical observations appeared to be the primary cause of Miras not having an optical counterpart. Thus, the group of gold Miras is a random sample of the bronze Miras, and not physically distinct.

The similarity of the zeropoints between the three samples in NGC 4258 suggests that having additional observations with a longer baseline or more frequent visits, while useful, is not essential to accurately determining the periodicity of objects in our sample. In addition, cross-matching previously known Cepheids to sources in our NGC 4258 master frame revealed that none of the known Cepheids in this field had been falsely detected as Miras. We were also unable to recover any Cepheids as variables based solely on their NIR variability without using prior knowledge on their locations. Some contamination by objects of the wrong period is unavoidable, but we estimate the number of contaminants to be most likely less than $\sim 5\%$ of our sample, and even smaller after outlier rejection.

\setlength{\tabcolsep}{1em}
\begin{deluxetable*}{lccc}
\tabletypesize{\scriptsize}
\tablecaption{Mira Sample Criteria}
\tablewidth{0pt}
\tablehead{ \colhead{} & \colhead{NGC 1559} & \colhead{NGC 4258 (gold)}
}

\startdata

    Period Cut (days): & $240 < P < 400$ &  $P < 300$  \\
    Amplitude Cut (mag): & $0.4 < \Delta \emph{F160W} < 0.8$ & $0.4 < \Delta \emph{F160W} < 0.8$ \\
    Surface Brightness Cut: & \surfbrightness counts/second &  --- \\
    $F$-statistic: & $\chi^2_{\text{s}}/\chi^2_{\text{l}} < 0.5$ & --- \\
    Color Cut (mag): & --- &$m_{F125W} - m_{F160W} < 1.3$  \\
    \emph{F814W} Detection: & --- & Slope-fit to \emph{F814W} data$ > 3\sigma$  \\
    \emph{F814W} Amplitude (mag): & --- & $\Delta\emph{F814W} > 0.3$ \\

\enddata
\label{tab:samples}
\tablecomments{A comparison of the criteria for the final Mira samples in NGC 1559 and NGC 4258. Owing to differences in signal-to-noise ratio and the available data, we were unable to match the criteria exactly.}
\end{deluxetable*}

\subsection{Recovery of Mira Parameters}
\label{sec:hamp}

Following H18, we consider only variables in the amplitude range $0.4 < A < 0.8$ mag to be candidate O-rich Miras. The minimum amplitude removes semiregular variables (SRVs) from our sample, which typically have smaller amplitudes \citep{Soszynski13}, or heavily-blended Mira stars, which will be artificially bright and have reduced amplitudes by their blending with a constant contaminant. The maximum amplitude is used to limit the number of heavily reddened C-rich Miras in our sample, which have on average larger amplitudes than O-rich Miras with the same period range. While C-rich Miras with low reddening track the O-rich Mira PLR, heavily-reddened C-rich Miras fall below the PLR and would bias our fit \citep{Whitelock03}. From our simulations, we find that our amplitudes are well-recovered, with the mean amplitude of the input and recovered Miras within a few 0.01\,mag of each other. 

\cite{Yuan17b} found larger amplitudes for C-rich Miras than O-rich Miras in the LMC and \cite{Cioni03b} in SMC variables. \citet{Goldman19} found that in IR wavelengths, both period and amplitude of pulsating AGB stars correlated with color, with redder variables having larger amplitudes. However, this effect does not appear in \cite{Soszynski09b}, where the LMC C-rich stars, which are typically redder than their O-rich counterparts, do not have larger amplitudes. It is possible that part of the discrepancy may lie in inconsistencies in the definition of amplitude (since the OGLE data is longer-baseline and we use only single-cycle amplitudes) or perhaps the range of pulsational amplitudes for each Mira spectral type differs between galaxies. The latter would imply that the amplitude cuts may need to be different in each galaxy. However, even in the metallicity range of the DUSTiNGS galaxies in \cite{Goldman19} there does not appear to be an obvious trend in galaxy metallicity and amplitude. As a result, we still use an amplitude cut and include possible effects of the amplitude cut on the zeropoint in our metallicity-related error budget (discussed in greater detail in \S \ref{sec:crich}). A more complete understanding of metallicity will be necessary to place firmer constraints on the systematic effect.

\begin{figure}
\epsscale{1.2}
  \plotone{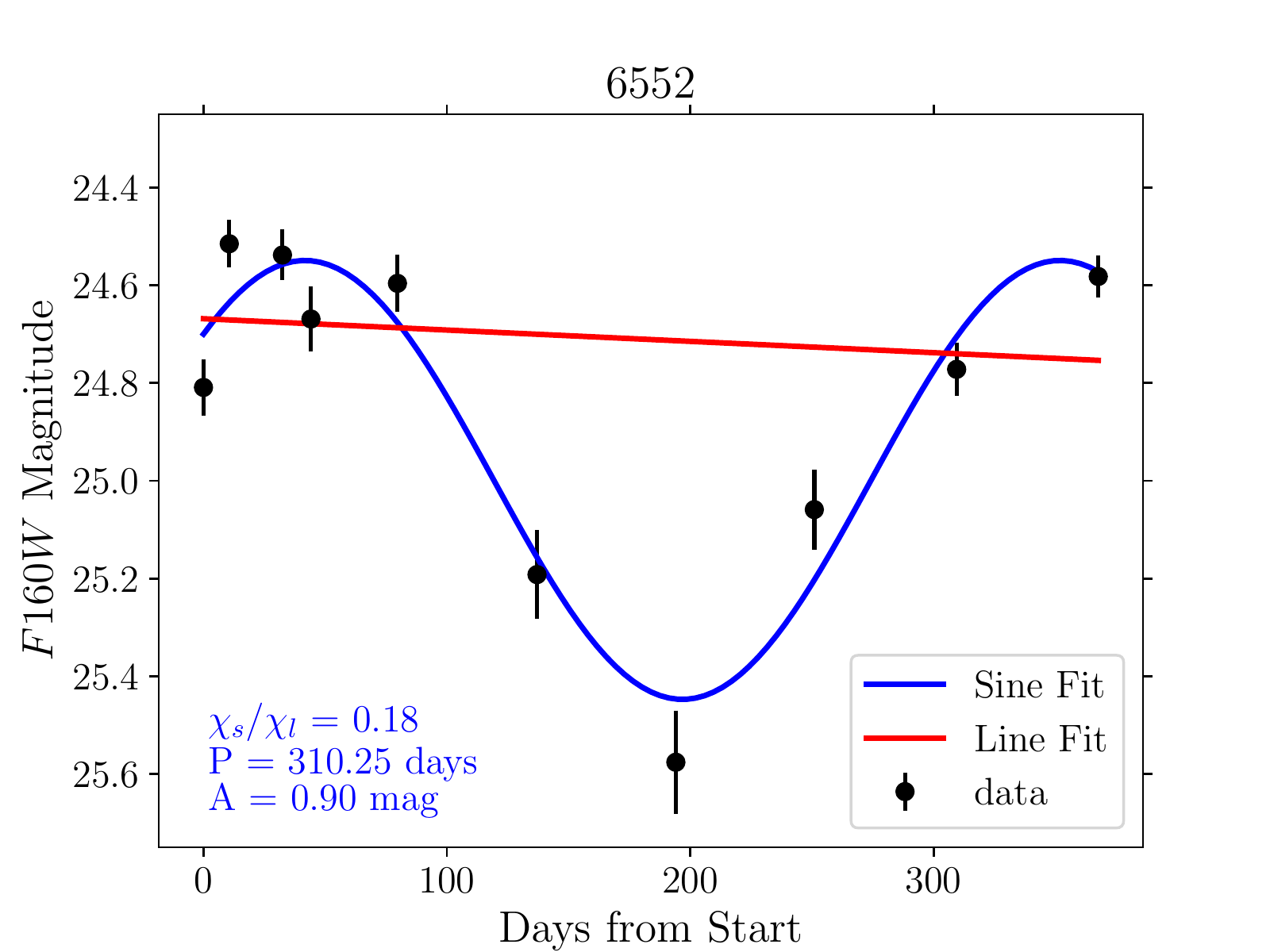}
  \plotone{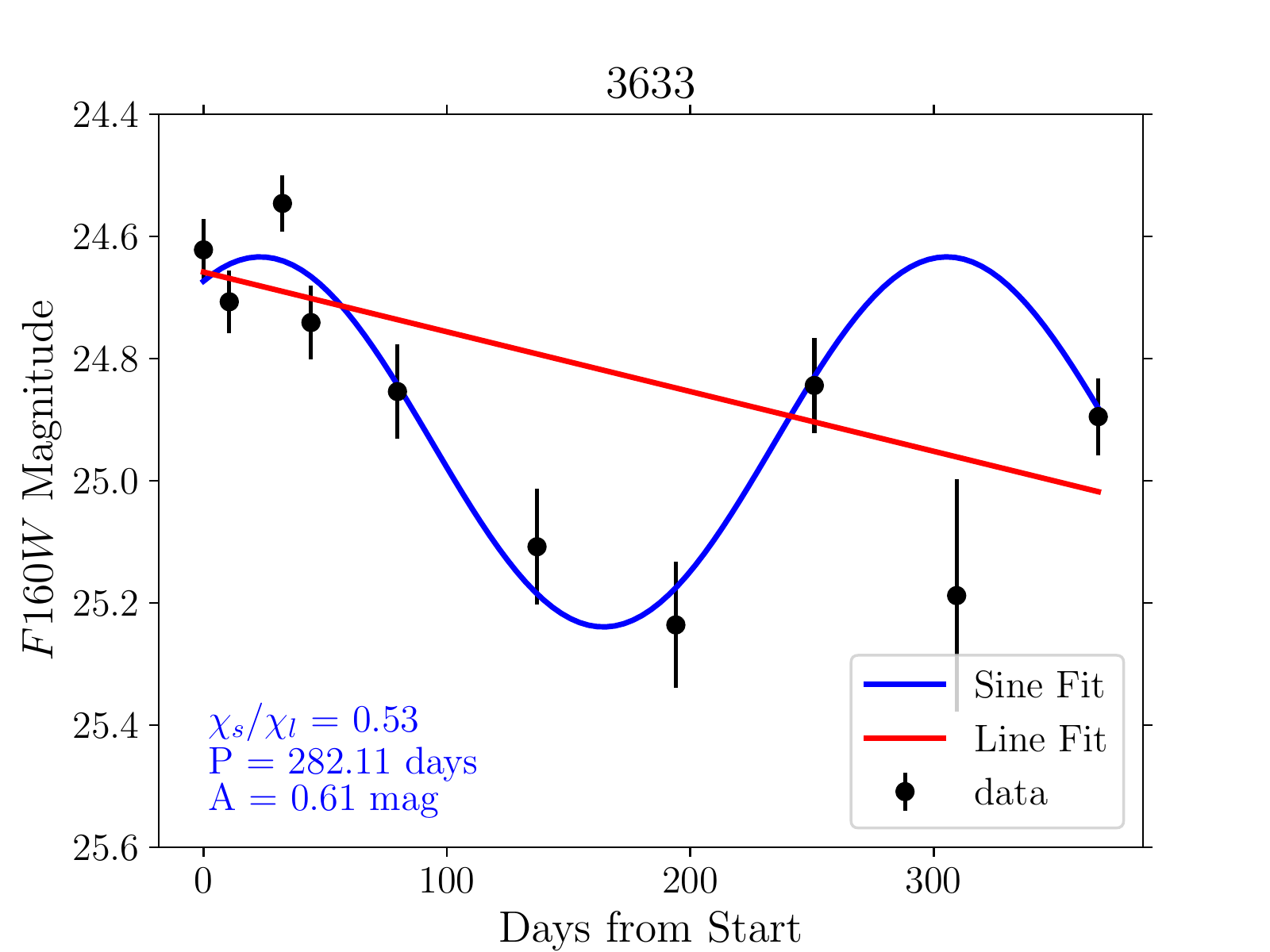}
  \caption{{\it Top:} An example of a variable that passed the $F$-statistic cut. {\it Bottom:} An example of a variable that failed the reduced $F$-statistic cut.
   }
  \label{fig:redchi}
\end{figure}

Besides using the period and amplitude of the sine fit, to remove imposters we may also compare the quality of the fit to that of a monotonic rise or fall. We compare each object's goodness of fit to a sine and a line using the $F$-statistic, which is defined as
\begin{align}
F = {\chi^2_s}/{\chi^2_l},
\end{align}
\noindent
where $\chi^2_s$ is the reduced $\chi^2$ of a sine fit and $\chi^2_l$ is the reduced $\chi^2$ for a line fit. This statistic may be used to retain only objects with light curves for which a sine fit is preferred over a straight line.
We use simulations to determine the best value for the $F$-statistic to retain Miras while removing noisy, static stars. We find that keeping sources with $F < 0.5$ is a fairly optimal choice; it removes $\sim 92$\% of the nonvariable stars while keeping $\sim 80$\% of all of our simulated variables. 

Examples of the straight-line and sine fits are shown in Figure \ref{fig:redchi}. We further visually inspect all of the objects that pass these criteria to remove spurious fits. 

\subsection{Period Cut}
\label{sec:percut}

\begin{figure}
\epsscale{1.2}
  \plotone{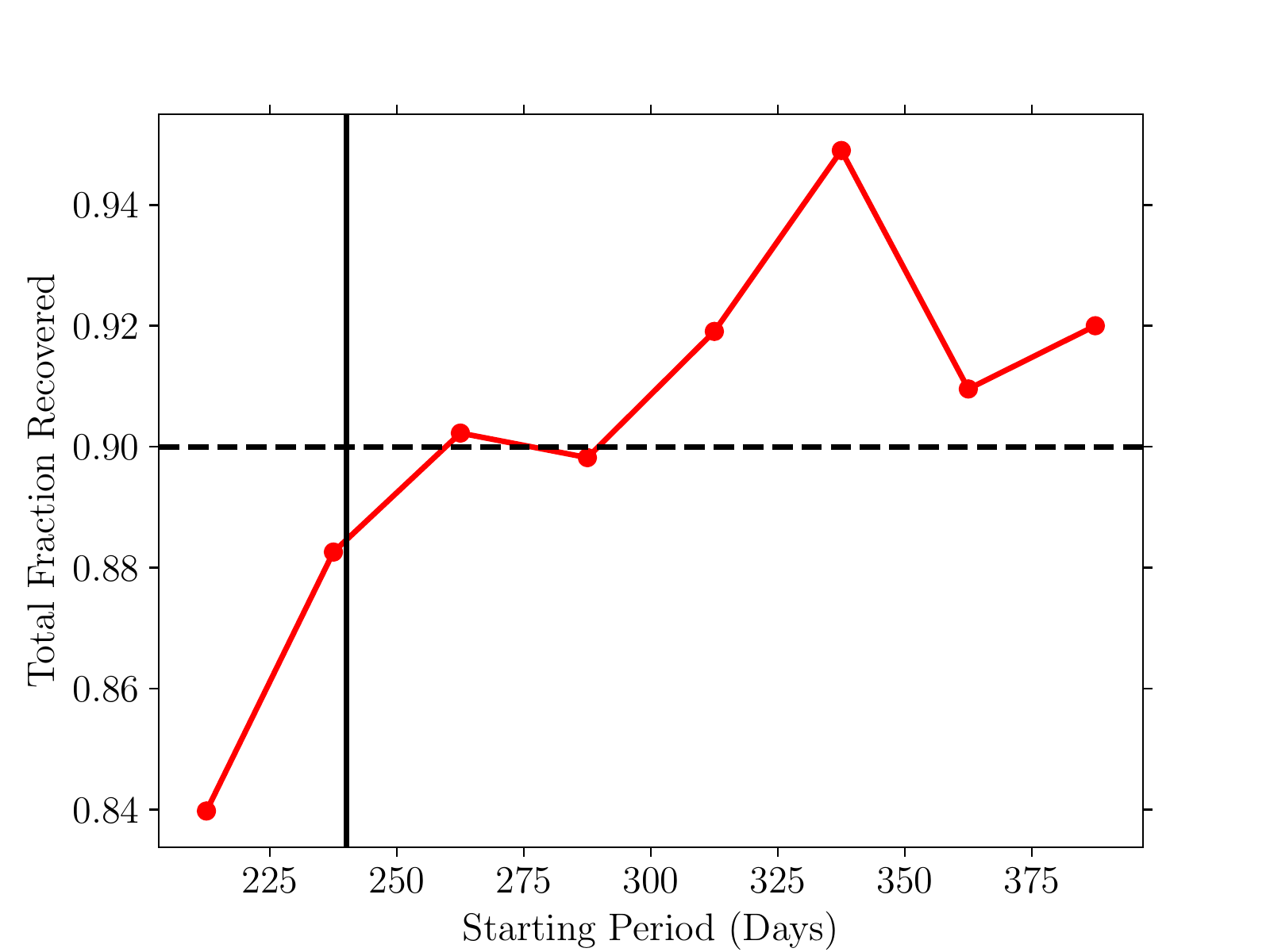}
  \caption{The fraction of simulated Miras recovered in our final sample as a function of period and smoothed into 25-day bins. We considered a period recovered if it was within 15\% of the true period. The simulated Miras that passed our variability criterion were considered to be recovered as variables. The dashed line shows the 90\% completeness limit. The black line shows the adopted 240-day lower period limit. 
   }
  \label{fig:recoveryrate}
\end{figure}

\begin{figure}
\epsscale{1.2}
  \plotone{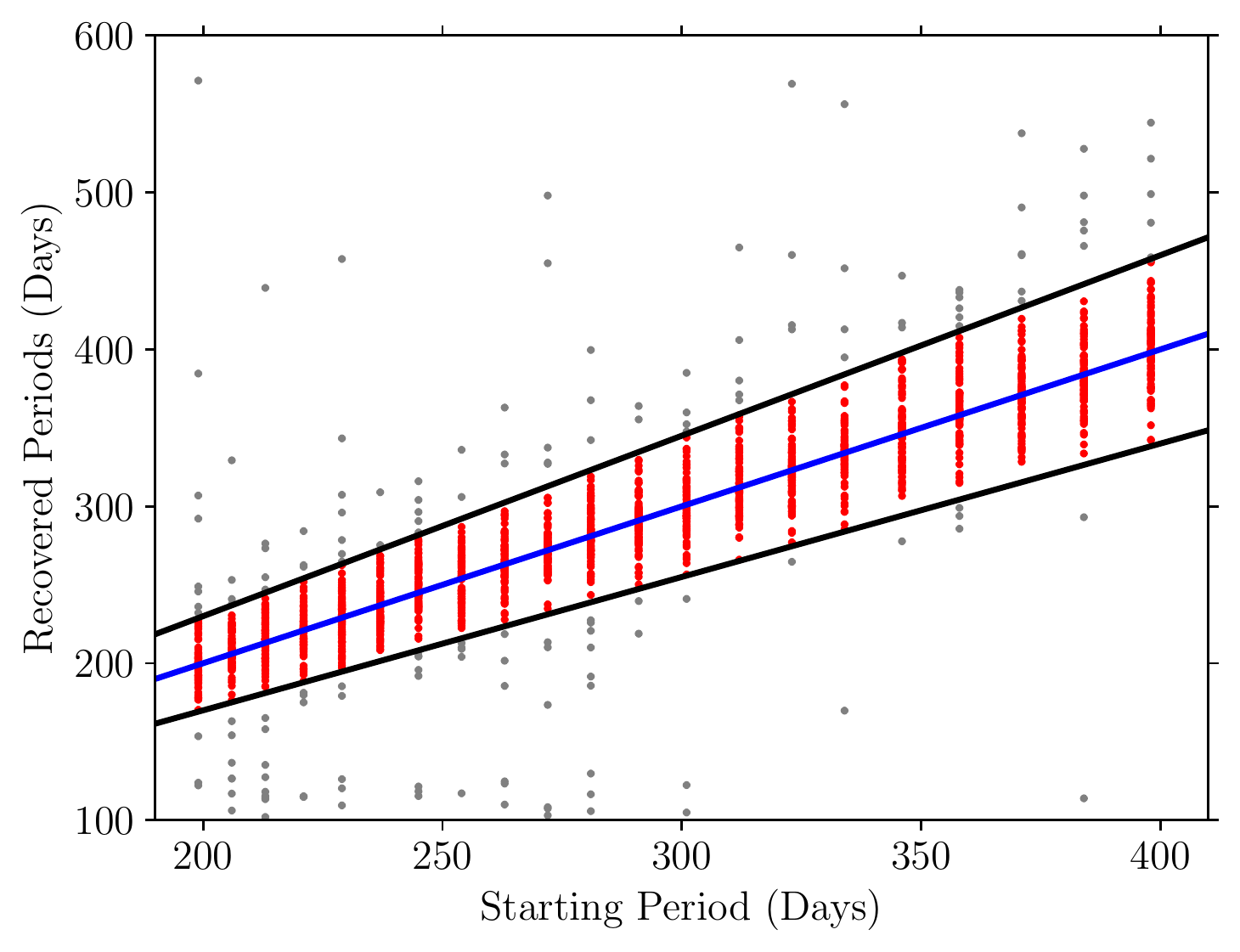}
    \caption{The input and recovered periods in our simulations. Red points are considered ``recovered" (within 15\% of the true input period). Black lines indicate the region of recovered periods. Gray points show periods that were not recovered by the simulation. The blue line shows $x = y$, where the true and recovered periods match exactly.
   }
  \label{fig:recoveryperiods}
\end{figure}

\begin{figure}
\epsscale{1.2}
  \plotone{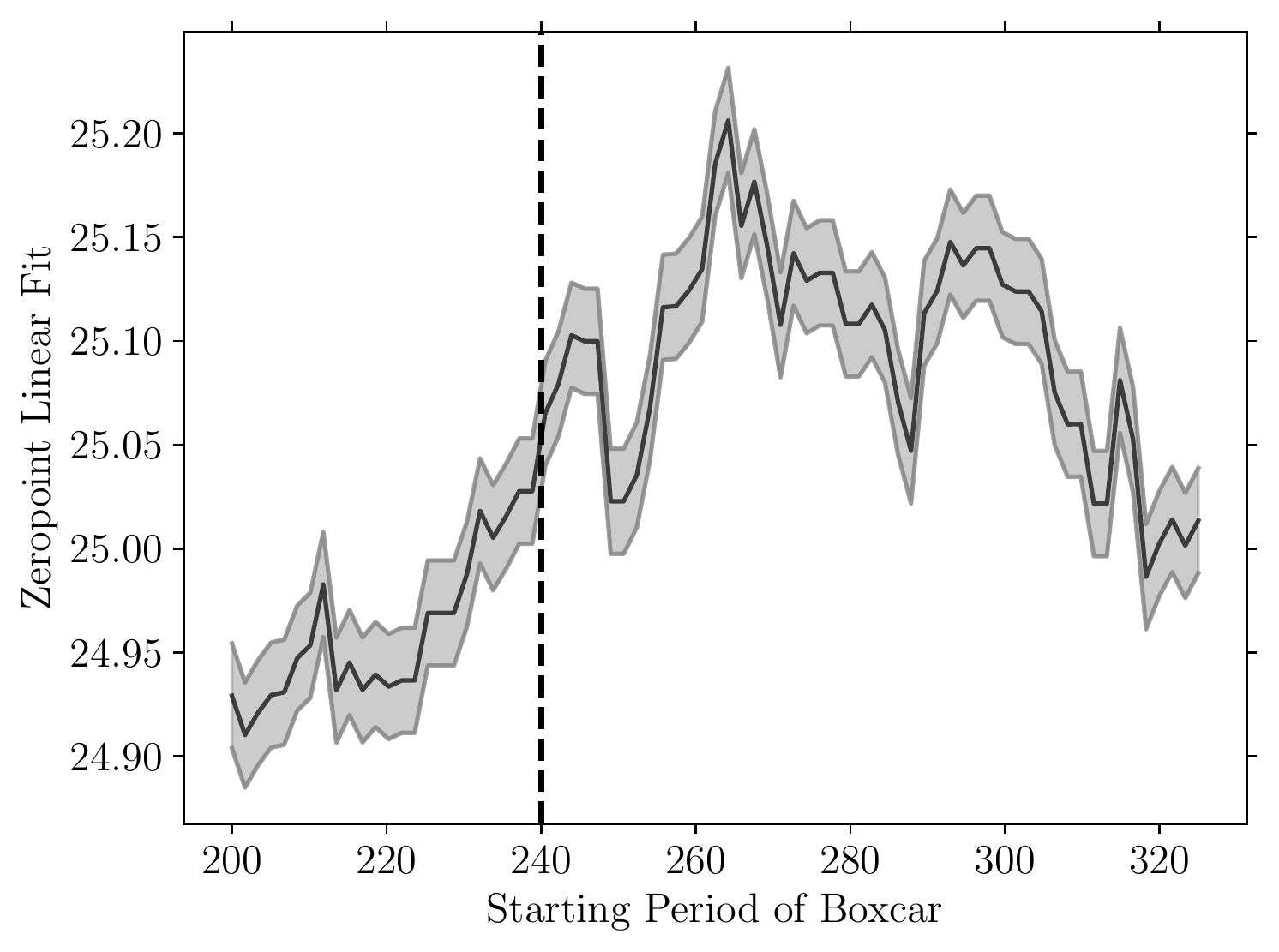}
  \caption{The zeropoint as a function of the starting period of each bin, using a boxcar fit with a width of 75 days. At $\sim 240$ days (black dashed line) the zeropoint starts to converge and oscillate about the true value. We chose 240 days as the minimum period based on this result. The black line shows the zeropoint using the \emph{F160W} slope. The gray lines denote the uncertainty in the zeropoint at each point. }
  \label{fig:boxcar}
\end{figure}

We limit our sample to $P<400$ days to exclude Miras undergoing hot-bottom-burning (although we would expect few if any HBB stars in this part of the galaxy because HBB stars typically have masses $> 3\, M_{\odot}$, which will be rare). An additional rationale for our period limit is that our observations span $\sim 370$ days, making the accurate recovery of periods greater than 400 days problematic. Figure \ref{fig:recoveryrate} illustrates the simulated period recovery rate as a function of period, and Figure \ref{fig:recoveryperiods} shows the recovered and input periods. We considered a period recovered if the measured and true values were within 15\% of each other. As a statistical error, this criterion would limit a period uncertainty to less than the apparent scatter of the PLR.  

We also limit our sample at the short-period, faint end of the PLR, owing to magnitude incompleteness. The Mira PLR has an intrinsic width and not all stars at the same period have the same magnitude. This means that the faintest stars at shorter periods are undetected, biasing us toward the brighter end of the PLR. Additional scatter due to the crowding and photometric errors further increases this magnitude range. We search for the minimum period above which we avoid this bias using three approaches.

First, to obtain an idea of the expected lower bound on our search for the completeness limit, we use the \emph{HST} Exposure Time Calculator (ETC) to calculate the estimated signal-to-noise ratio (SNR) for each period range. We derive a preliminary fit to the candidate Mira PLR to relate Mira periods to their mean magnitudes.  We then determine whether that mean magnitude would be detected at SNR $>5$ at the depth of each imaging epoch, a minimum requirement for useful detection.  This analysis leads us to draw the initial lower bound at $\sim 200$ days. While this is the theoretical low-period limit to which we should discover Miras, in practice, crowding and the intrinsic width of the Mira PLR introduce additional scatter causing some Miras near that limit to go undetected.  Thus, we only use the ETC results as a starting point. 

Next, we proceed to look for the expected empirical signature of incompleteness --- an approximate plateau of the PLR intercept or zeropoint at long periods and a brightening trend of the zeropoint as we cross the period completeness limit toward the short-period direction. In order to find the location of this break, we start at the initial value of 200 days from the \emph{HST} ETC analysis and use a moving 75-day window (boxcar) to reduce the noise and measure the period-range average of the zeropoint. The result is shown in Figure \ref{fig:boxcar}. At $P>240$ days the trend reaches a plateau (with oscillations due to noise and irregular sampling above this). 
 
Finally, we use simulations (described in Section \ref{sec:sims}) to verify this empirical result based on the {\it expected} recovery rate. 

Figure \ref{fig:recoveryrate} displays the number of sources recovered as a function of starting period. We can see that an increasing fraction of stars is recovered as the input period increases, and at $\sim 240$ days the recovery fraction reaches $\sim 90$\%, which is in accord with the empirical analysis. Therefore, we choose this period as our short-period end for the completeness. 

\subsection{Surface Brightness and Crowding}
\label{sec:surfbrightness}

\begin{figure}
\epsscale{1.2}
  \plotone{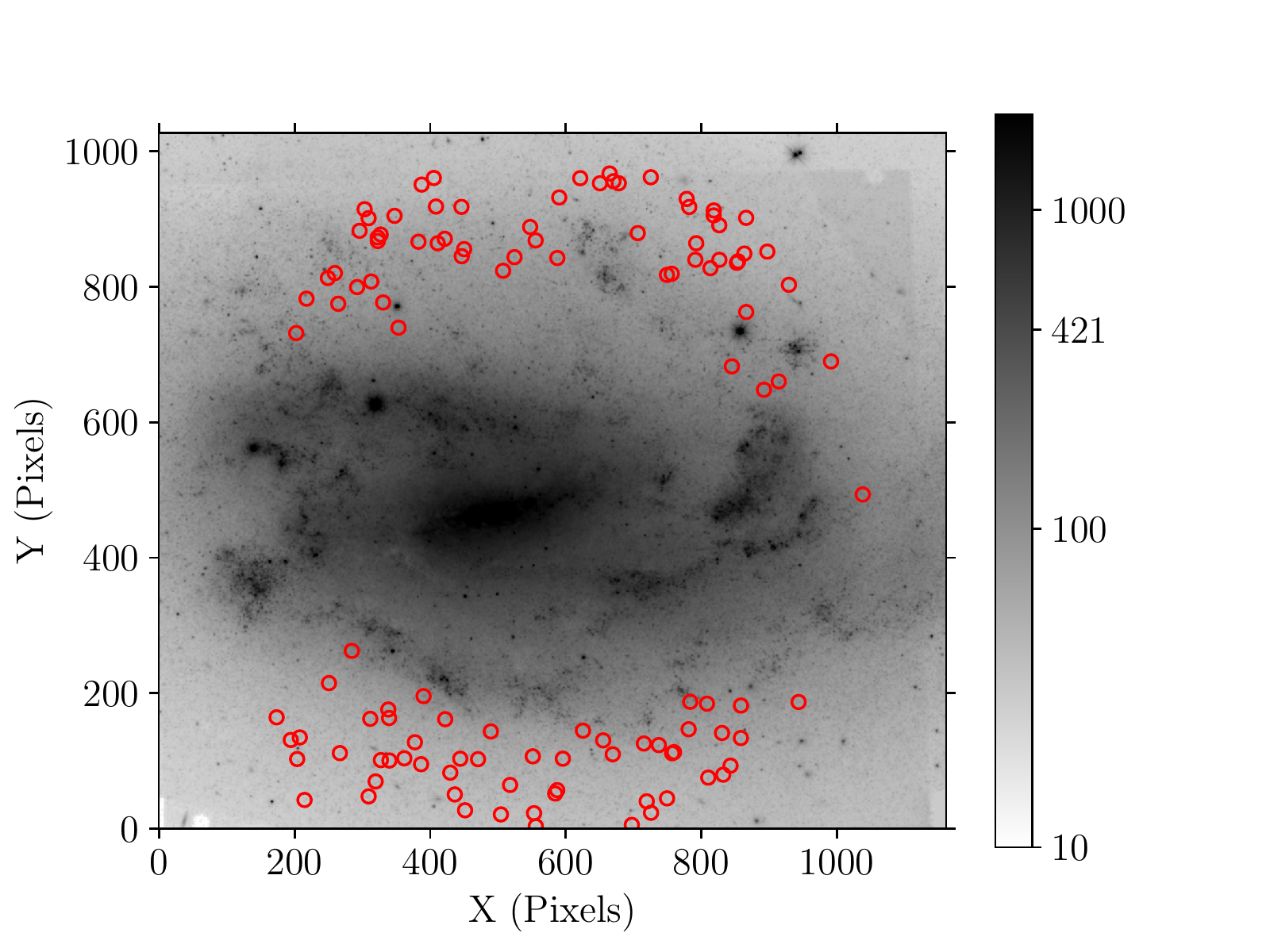}
  \caption{The distribution of Miras (in red circles) overplotted on the \emph{F160W} master image. The colorbar shows the surface brightness in counts arcsec$^{-2}$. We used a surface brightness cutoff of 421 counts s$^{-1}$ arcsec$^{-2}$, indicated on the colorbar.
   }
  \label{fig:miraloc}
\end{figure}

Cepheid analyses \citep{Riess16, Hoffmann16} in the NIR routinely use artificial star injection and recovery to correct for the artificial brightness in detected sources due to unresolved background sources --- that is, crowding. We follow this approach. Like the Cepheid analyses, we choose not to include objects in very crowded regions, thereby reducing the size of this correction (and its associated uncertainties).  In the case of NGC 1559, this is most effectively addressed by including a requirement for low surface brightness background.  

NGC 1559 has high surface brightness near the center, causing the image to be particularly crowded toward the nucleus of the galaxy (see Figure \ref{fig:miraloc}). The inclination of the galaxy, $i \approx 57.3^\circ$ \citep{Kassin06}, is a likely contributing factor to its high surface brightness near the center.   Through our artificial star tests (described in the following section), we found that the crowding bias corrections would generally exceed 1\,mag throughout an elliptical region around the center, meaning that these sources are dominated by the background, and will lose reliability compared to Miras in less crowded regions. We measured the local surface brightness in this region as the average number of counts per second in a 2$''$-wide box around each candidate. We identified  a minimum surface brightness of $\sim 420$ counts per second per square arcsecond as corresponding to this contiguous region and set this as the boundary for the maximum allowed surface brightness to be included in our sample.   This leaves objects that are mainly distributed in the outer regions of the galaxy, as seen in Figure \ref{fig:miraloc}.   As expected of aged populations, the Miras in NGC 1559 are evenly distributed in the regions of lower surface brightness, and do not follow spiral arms or other galactic structures.

\begin{figure}
\epsscale{1.2}
  \plotone{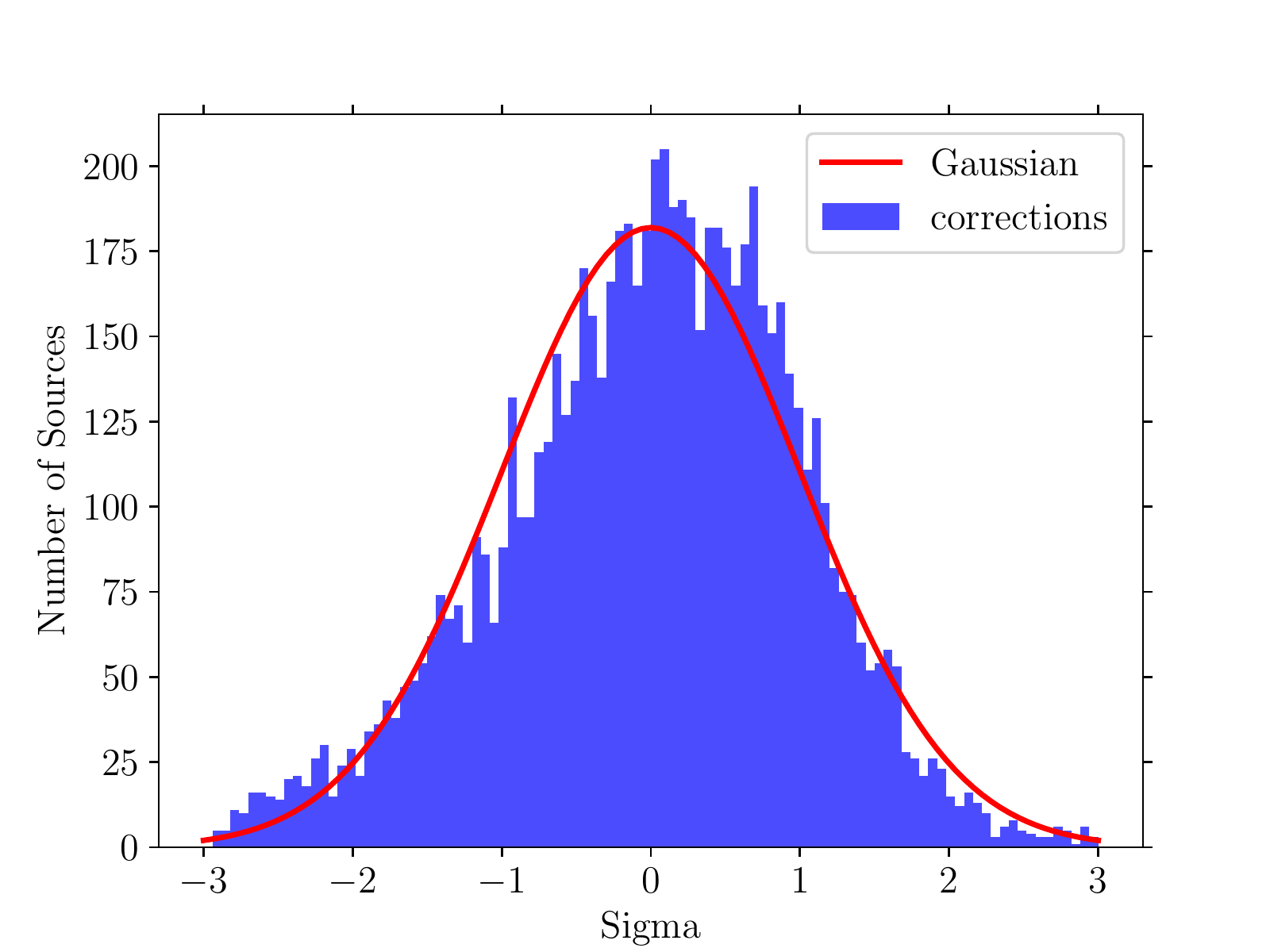}
  \caption{ The distribution of crowding corrections for all of the artificial stars, after $3\sigma$ clipping and removing blended artificial stars. For each Mira, we use 100 artificial stars to measure its crowding correction. We then subtract the mean crowding correction for each Mira from its set of 100 artificial stars and divide all of the corrections by the standard deviation. The red line shows a Gaussian distribution, while the blue histogram is the actual distribution.  }
  \label{fig:crowding}
\end{figure}

After using surface brightness cuts to remove objects that would be in the most crowded regions, we employ artificial star tests to determine the individual crowding bias for the remaining candidate Miras. For each remaining Mira candidate we fit an initial PLR. We then inject 100 artificial stars for every Mira, using the Mira's period and the initial PLR fit to determine a starting point for the artificial star magnitude. The artificial star is randomly dropped within a 2$''$ radius of the Mira, so that both are in the same environment. 

We perform aperture and PSF photometry on the artificial stars exactly as we would with the real Miras and measure the mean magnitudes at which they are recovered. The difference in magnitude between the input artificial stars and the output is used to get a first estimate of the correction for each star. The magnitude of the original Miras is then corrected before refitting the PLR to determine a new guess for the true magnitudes, and the process is repeated until the Mira magnitudes converge. The corrections apply to our use of the mean magnitude of the Mira (defined by the mean of the sine fit). 

Occasionally, our artificial stars will fall close enough to a bright background star that the two will be``blended" (i.e., unresolved). If the superimposed star is sufficiently bright, this blending would suppress the amplitude of a variable star enough that it is unlikely to be recovered as a Mira based on our minimum-amplitude requirement. We approximate the blended amplitude by the expression 
\begin{equation}
F = 10^{-0.4 \Delta m} + 10^{-0.4A\phi},
\end{equation}
where $\Delta m$ is the difference in magnitude between the two stars (positive if the background star is fainter), $A$ is the amplitude of a Mira, and $\phi$ is the phase of the Mira. We can likewise calculate the magnitudes at the peak and trough of the lightcurve and subtract the two to obtain a new blended amplitude. At an original amplitude of 0.6 mag (mean of our range), and a background star 1 mag fainter than the artificial star, we find that the blended amplitude is $\sim 0.4$ mag, our cutoff for the low-amplitude end. Blended stars brighter than this threshold would cause the mean Mira  to be removed solely on the basis of amplitude cuts. Thus, we consider the artificial star to be blended if it is within 1 pixel of a bright star (defined as $ < 1$ mag fainter than the Mira) and these (like real blended Miras) are not included in our artificial star samples. 

After a $3\sigma$ clip of the artificial star sample (the same clip we eventually use when fitting the PLR), we take the standard deviation of these crowding corrections to get an estimate for the uncertainty in each Mira's mean magnitude. We use the mean of the clipped crowding corrections to correct the magnitude of each Mira. Figure \ref{fig:crowding} shows the distribution of crowding corrections for all of our artificial stars, after subtracting the mean correction for each Mira and dividing by the standard deviation of the corrections for each Mira. This distribution in magnitudes compares well to a Gaussian distribution (in red).  

We also conservatively eliminate 5 Mira candidates (3\% of the sample) that have a crowding correction $\Delta m_b > 0.5$ mag. Although these may yield good measurements, this level of crowding is significantly greater than for the rest of the sample (the mean crowding correction is $\sim 0.13$ mag, with a standard deviation of 0.14 mag). In addition, applying a crowding correction reduces the scatter  of the PLR by approximately 20\%. The local surface brightness criterion already excludes the majority of candidates that would have fallen into this category, so this final cut ensures that all of the objects remaining are in uncrowded regions. 

\subsection{C-Rich Mira Contamination}
\label{sec:crich}

\begin{figure}
\epsscale{1.2}
  \plotone{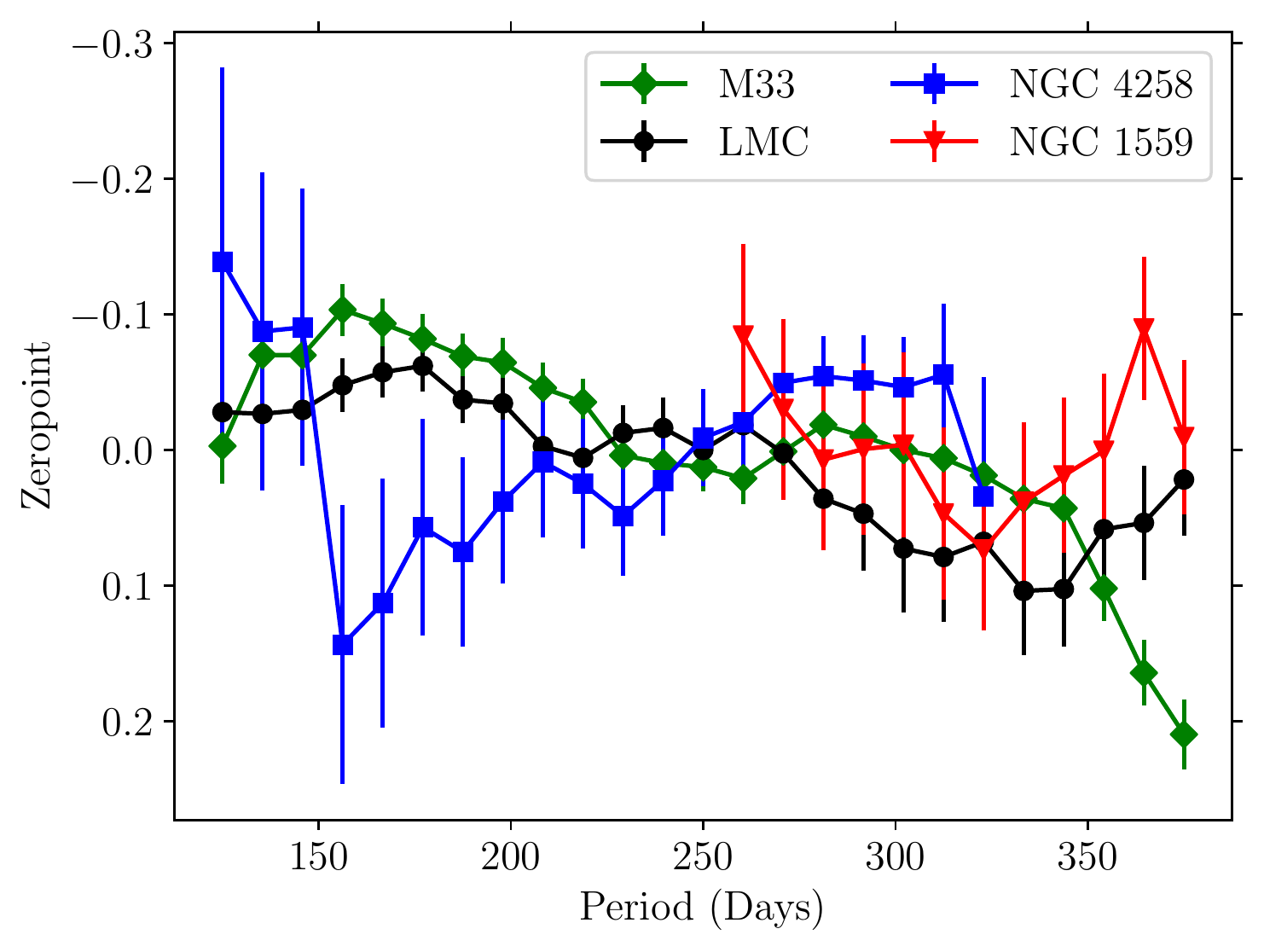}
  \caption{The zeropoint as a function of period for the contaminated samples from the LMC and M33 (in black and green, respectively) and for NGC 1559 and NGC 4258 (in red and blue, respectively). For the LMC and M33, we first convert the ground-based $J$ and $H$ data to \emph{F160W} using our color term and then apply an amplitude cut to the combined C-rich and O-rich Mira dataset to simulate a contaminated sample before fitting the zeropoint in each bin.}
  \label{fig:zptall}
\end{figure}

Although we used amplitude cuts to reduce the number of C-rich Miras or nonvariable stars in our sample, here we quantify and correct for a potentially low level of residual contamination from C-rich Miras.  C-rich Miras are typically fainter than O-rich Miras in \emph{F160W}; thus, C-rich contamination in the Mira PLR can result in a fainter absolute magnitude calibration. We employ the OGLE-LMC sample --- the most complete sample of Mira variables with well-categorized Miras --- to measure the difference in zeropoint between a pure O-rich sample and a model C-rich-contaminated sample (by mixing O-rich and C-rich LMC Miras).  We note that this is not a large correction because for the LMC, the contaminated sample would shift the zeropoint by $\sim 0.07$ mag.  

To create the LMC ``contaminated sample," we convert the 2MASS ground-based $J$ and $H$ data from the LMC to \emph{F160W} using Equation \ref{eq:colorcorr} and then apply the same amplitude cuts and sigma-clipping used for NGC 4258 and NGC 1559.  The single-epoch amplitudes in the $H$ band are estimated from the $H$-band light curve templates from \cite{Yuan17b}. In the LMC, at $P<400$ days, the ratio of O-rich to C-rich Miras is approximately $2:3$ before the amplitude cuts and $2:1$ after.   As expected, the amplitude cuts remove the majority of C-rich Miras in this range. Then we calculate the zeropoint (i.e., the PLR intercept) as a function of period by fitting a PLR with a fixed slope to the Mira sample binned by 50 days. As shown in Figure \ref{fig:zptall}, the zeropoint of the LMC contaminated sample grows fainter as the contamination by C-rich Miras increases ($P \gtrsim 250$ days), whereas a purely O-rich sample ($P < 250$ days) is relatively flat.  Therefore, we can use the change in zeropoint as a function of period of the LMC contaminated sample as a template to measure the contamination in the other hosts. After plotting the zeropoint as a function of period for each galaxy, we use the amplitude and shape of the curve for each galaxy relative to the amplitude and shape LMC curve in order to estimate the fraction of C-rich contamination is present in that galaxy's Mira sample.

We take the samples of Miras in M33 (also well-categorized into C-rich and O-rich subclasses) from \citet{Yuan18} as a test of our ability to measure the contamination.  The observations of M33, like NGC 1559 and NGC 4258, impose a magnitude limit on the Mira sample such that the C-rich population is relatively incomplete compared to the O-rich population.  We combine the O-rich and C-rich samples from \citet{Yuan18} which pass our amplitude cuts and measure the zeropoint as a function of period. Adopting the LMC curve from Figure \ref{fig:zptall} as a model for C-rich contamination, we fit the M33 curve from Figure \ref{fig:zptall} using the form
\begin{align}
Z_{1} = \alpha Z_{\text{LMC}} + \beta,
\end{align}
where $Z_{1}$ is the zeropoint of the contaminated M33 PLR as a function of period, $Z_{\text{LMC}}$ is the zeropoint of the contaminated LMC PLR as a function of period, and $\alpha$ and $\beta$ are fit parameters: $\beta$ simply shifts the curve and does not contribute to a correction, while $\alpha$ tells us the amount to correct the zeropoint as a function of the total difference in zeropoint between the combined and O-rich-only values.

For M33 we find that $\alpha = 0.58 \pm 0.18$ is the best fit to the LMC curve.  For the LMC, the difference in zeropoint is $-0.07 \pm 0.013$ mag. Therefore, in order to correct the zeropoint of the combined C-rich and O-rich M33 dataset to match the zeropoint of the O-rich-only M33 set, $m_c = -0.04 \pm 0.01$ mag. The known difference in zeropoint of the purely O-rich and the contaminated PLR is $-0.034$ mag, within the uncertainties of our contamination correction. We now apply this same technique to NGC 1559 and NGC 4258.

In Figure \ref{fig:zptall}, we plot the change in zeropoint as a function of period for NGC 4258 and NGC 1559.  Applying the LMC as a model to NGC 4258 and NGC 1559, we find that $\alpha = -0.77 \pm 0.25$ for NGC 4258, but $\alpha = 0.81 \pm 0.35$ for NGC 1559, indicating some possible C-rich contamination. We determine an overall correction to the zeropoint of NGC 1559 of $m_c = -0.057 \pm 0.024$ mag.  We apply this correction to the $m_f$ given in Eq. \ref{eq:mag},
\begin{equation}
\label{eq:contaminationmag}
m_{cf} = m_f + m_c,
\end{equation}
where $m_c$ is the C-rich correction. For NGC 4258, where $\alpha < 0$, we do not correct for any possible C-rich contamination, since C-rich contamination should only make the zeropoint fainter, not brighter. 

We would expect to find more C-rich contamination in NGC 1559 than in NGC 4258 because the range of Mira periods extends up to 400 days in NGC 1559 and only up to 300 days in NGC 4258, and there are more C-rich stars at longer periods. Depending on the amount of circumstellar extinction, C-rich Miras can be as much as 2--3 mag fainter in the $H$ band than O-rich Miras having the same period \citep{Ita11}. Thus, we would also expect to find less contamination in magnitude-limited samples like NGC 1559 and NGC 4258 than in the complete sample in the LMC, in good agreement with the results of our C-rich contamination corrections.

\subsection{Simulations}
\label{sec:sims}

To test the efficacy of our selection criteria, we use simulated light curves of Miras and nonvariable stars. We also use the simulations to determine the amount of contamination that we might expect from nonvariable stars, and the completeness limit of the sample. For each simulation, we inject 105 artificial variable stars into the star catalogs at 21 periods between 200 and 400 days. The artificial stars have random phases, similar to our Mira sample. We use a sine wave of the appropriate period and a starting amplitude of 0.7 mag, the estimated median ``true" amplitude of our sources. We then include a scatter of 0.4 mag to the mean magnitude of the source to mimic the spread of our Mira PLR. Next, using the \emph{HST} ETC, we estimate the SNR of each source, add in the appropriate noise, randomize the starting phase, and sample the light curve at the observation dates of our survey. We perform the rest of the variability search on these stars just as we would for the true Miras. The same is done for nonvariable stars to see the effects of the variability cuts on both the simulated Miras and nonvariable stars. Nonvariable stars are simulated by using starting magnitudes in the same range as the Mira variables, but with zero amplitude. 

To test the level of nonvariable star contamination, we simulate 10,000 nonvariable stars at the catalog level and then apply all of our Mira selection criteria (outlined in Sections \ref{sec:variability}, \ref{sec:hamp}, and \ref{sec:percut}). We find four stars that passed all of the variability cuts. However, all four stars also fall more than 1.5 mag below the Mira PLR and were sigma-clipped out of the final sample. Thus, we conservatively estimate that we may have 1 nonvariable star contaminant per 10,000 nonvariable stars in the master star list. Out of the observed star list of 49,000, approximately 40,000 are nonvariable stars. We expect an upper limit of four nonvariable star contaminants in the Mira PLR and we do not expect these to meaningfully bias the PLR.  

\section{Systematics}
\label{sec:sys}

In this section, we discuss the sources of systematic uncertainties in our analysis, summarized in Table 5. Specifically, we consider the effect of slope on the zeropoint (\S \ref{sec:sysslope}) and potential systematics from reddening (\S \ref{sec:reddening}). Our present uncertainty in $H_0$ is dominated by the uncertainty from a single SN~Ia, with the systematic uncertainties discussed below being subdominant.  

\setlength{\tabcolsep}{1em}
\begin{deluxetable*}{lcc}
\tabletypesize{\scriptsize}
\tablecaption{Systematic and Statistical Uncertainties}
\tablewidth{0pt}
\tablehead{\colhead{Source} & \colhead{Systematic Uncertainty} & \colhead{Statistical Uncertainty}\\
\colhead{} & \colhead{(mag)} & \colhead{(mag)}
}
\startdata

$a_B$ & 0.00176 & --- \\
Aperture Correction & --- & 0.01 \\
C-rich Correction & --- & 0.024 \\
Color Term & 0.02 & --- \\
Differential Extinction & \diffreddening & --- \\
LMC Distance Modulus & 0.0263 & --- \\
LMC Zeropoint & --- & 0.01 \\ 
Metallicity & 0.03 & --- \\
NGC 1559 Zeropoint & --- & 0.038 \\
NGC 4258 PLR & --- & 0.017 \\
NGC 4258 Distance Modulus & 0.032 &  --- \\
Uncertainty in Slope (both slopes) & \slopeuncertainty & --- \\
Supernova Peak Magnitude & --- & 0.11 \\ & &\vspace{-.07in} \\
\hline \\
 Subtotals (NGC 4258) &  0.060 & \, 0.050   \\
\vspace{.07in}Subtotals (LMC) &  0.046 & 0.047
\enddata
\tablecomments{With the new maser distance to NGC 4258 \citep{Reid19}, the uncertainty in the extinction is now the dominant source of systematic uncertainty. Uncertainties shown are for preferred value of $H_0$ using two anchors and the slope we fit after converting ground-based data to \emph{F160W} ($-3.35$). The subtotals give the amount of statistial and systematic error for the calibration of the distance modulus of NGC 1559 ($\sigma_{\mu 1559}$ from Equation \ref{eq:Msig}) assuming a 240-400 day period range. The color-term uncertainty comes from H18, and is the estimated uncertainty from using a ground-based sample (LMC) to calibrate the \emph{HST} samples. Thus, it does not contribute to the systematic error when using the NGC 4258 only as an anchor. } 
\label{tab:sys}
\end{deluxetable*}

\subsection{Slope}
\label{sec:sysslope}

Miras and other variable stars are typically fit with a linear PLR, and we apply two different choices of slopes derived from LMC Miras.  Because the Mira samples in NGC 4258 and NGC 1559 have lower SNR and a shorter period range than those in the LMC, we do not attempt to use these to derive an independent slope, though this may become possible as the SN-Mira host sample grows.

The first slope is derived by \citet{Yuan17b}  using observations of about 170 LMC O-rich Miras in the $H$ band, which is the closest match to \emph{F160W}:
\begin{align} \label{eq:plrh}
m = a_0 - 3.64(\log P - 2.3),
\end{align}
where $P$ is the period in days, $m$ is the $H$-band magnitude, and $a_0$ is the intercept or zeropoint. For the ``gold'' sample of NGC 4258 Miras, we obtain $a_0 = \pone$ mag (statistical error only) with this slope. 

In addition to the slope determined by \citet{Yuan17b}, we use the OGLE LMC variable-star database to estimate the slope of the Mira PLR in \emph{F160W} by transforming from ground-based IR data to the \emph{HST} bandpass. Using 2MASS \emph{J} and \emph{H} data for OGLE LMC Mira variables, we adopt the color correction derived by H18, 
\begin{align} \label{eq:colorcorr}
m_{F160W} = H + 0.39(J-H),
\end{align}
to convert their O-rich Mira sample from ground-based $J$ and $H$ into \emph{F160W} magnitudes. 
This sample includes only periods below the HBB cutoff, $P=400$ days, to better match our \emph{HST} sample. Using this sample of 416 Miras with $P<400$ days, we obtain a slope fit of 
\begin{align} \label{eq:plrf160}
m = a_0 - 3.35(\log P - 2.3).
\end{align}
For this slope, we obtain $a_0 = \ptwo $ mag (statistical error only) for the NGC 4258 gold sample of Miras.  In Table 5 we include a systematic uncertainty based on the differences in $a_0$ when using these two slopes.

\subsection{Extinction and Metallicity}
\label{sec:reddening}

Using the \citet{Schlafly11} dust maps, we find that the Galactic extinction in the direction of NGC 1559 in the 2MASS \emph{J} band is $\sim 0.022$ mag and the 2MASS \emph{H} band is $\sim 0.014$ mag. However, for the purposes of calibration, we are affected by the difference in extinction between NGC 4258 and NGC 1559. The Galactic foreground extinction at the location of NGC 4258 in the 2MASS \emph{J} and \emph{H} bands is 0.012 and 0.007 mag (respectively), for a difference of 0.01 and 0.007 mag between the two, for which we correct as follows:
\begin{equation}
\label{eq:extinctioneq}
m_{\text{PLR}} = m_{cf} + m_e,
\end{equation}
where $m_{\text{PLR}}$ is the final mean magnitude of the Mira (which we use for fitting the PLR), $m_{cf}$ is the C-rich contamination-corrected mean magnitude defined in Equation \ref{eq:contaminationmag}, and $m_e$ is the correction for the differential foreground extinction, which we estimate to be 0.01 mag. 
The \emph{F160W} lies between the \emph{J} and \emph{H} bands, so we derive its extinction by interpolating between the \emph{J} and \emph{H} bands.

We expect the {\it difference} between the Mira interstellar extinction in NGC 1559 and NGC 4258 to be quite low.  Miras are not a disk population where dust is concentrated.  The positional distribution of the NGC 4258 Mira sample (see H18) is uniform, and thus should mimic a lower extinction halo distribution.  Likewise for NGC 1559, most of our sources are in the outer regions of the galaxy and thus would have low extinction.  Most importantly, our measurements are at 1.6\,$\mu$m which is much less sensitive to extinction than shorter wavelengths. 
However, there could be a small difference in the extinction between the regions where the NGC 4258 and NGC 1559 Miras are located. \citet{Riess09} suggested a $\sim 0.04$ mag differential extinction between Cepheids in the $H$ band between NGC 4258 and SN hosts, and we propagate this value as a systematic uncertainty to account for a difference in interstellar extinction.

While the effects of metallicity on the Mira PLR in the NIR are not well-studied, the NIR and IR wavelengths are known to be less sensitive to metallicity than the optical. In addition, there is evidence to suggest that the metallicity effect in NIR wavelengths is $< 0.1$ mag.  \cite{Goldman19} find that at IR wavelengths, the pulsation properties of AGB stars show no dependence on metallicity over a range of [Fe/H] = -1.85 (1.4\% solar) to [Fe/H] = -0.38 (50\% solar). \cite{Bhardwaj19} found (without correcting for metallicity) that the mean relative distance between the LMC and the SMC measured in the $K_s$ band using the O-rich Mira PLR to be consistent with the measurements from previous studies using Miras and Cepheids.

Finally, \cite{Whitelock08} found an upper limit of $\sim 0.1$ mag on difference in zeropoint between the LMC and Galactic O-rich Miras that have both Hipparcos parallaxes and VLBI parallaxes and Miras in Globular Clusters. However, this upper limit uses a distance modulus of $\mu = 18.39 \pm 0.05$ for the LMC distance. Current best estimates of the distance to the LMC \citep{Pietrzynski19} list this distance as $\mu = 18.477 \pm 0.004 \text{ (stat)} \pm 0.026 \text{ (sys)}$. With this revised LMC distance, the two zeropoints agree to within $\sim 0.01$ mag. In addition, NGC 4258 and NGC 1559 are both large spiral galaxies, and we expect them to have similar metallicities, further reducing the effects of metallicity on the PLR. However, we conservatively budget a $\sim 0.03$ mag systematic uncertainty on the zeropoint due to metallicity effects to account for this and potential amplitude differences (discussed in \S \ref{sec:hamp}).

\section{Results and Discussion}
\label{sec:results}

\setlength{\tabcolsep}{.7em}
\begin{deluxetable*}{lccc}
\tabletypesize{\scriptsize}
\tablecaption{Summary of Mira $H_0$ Values}
\tablewidth{0pt}
\tablehead{\colhead{Anchor} & \colhead{Period Range (days)} & \multicolumn{2}{c}{$H_0$ (km\,s$^{-1}$\,Mpc$^{-1}$)} \\
\colhead{} & \colhead{} & \colhead{Slope $= -3.35$} & \colhead{Slope $= -3.64$}}

\startdata

NGC 4258 & $240 < P < 300$ & $\HCpartp $& $\HCpartptwo$ \\
NGC 4258 & $240 < P < 400$  & $\HCfullp$ & $\HCfullptwo$ \\
LMC & $240 < P < 400$ & $\HCLMC$ & $\HCLMCtwo$ \\
{\bf LMC $+$ NGC 4258} & $240 < P < 400$ & {\bf 73.3} \boldmath$\pm 4.0 $ & $\HCCombinedtwo$ 
\enddata
\tablecomments{The $H_0$ with two anchors and the slope fit using \emph{F160W} ($-3.35$) is our preferred value.} 
\label{tab:sys}
\end{deluxetable*}

\begin{figure*}
\epsscale{1.2}
  \plotone{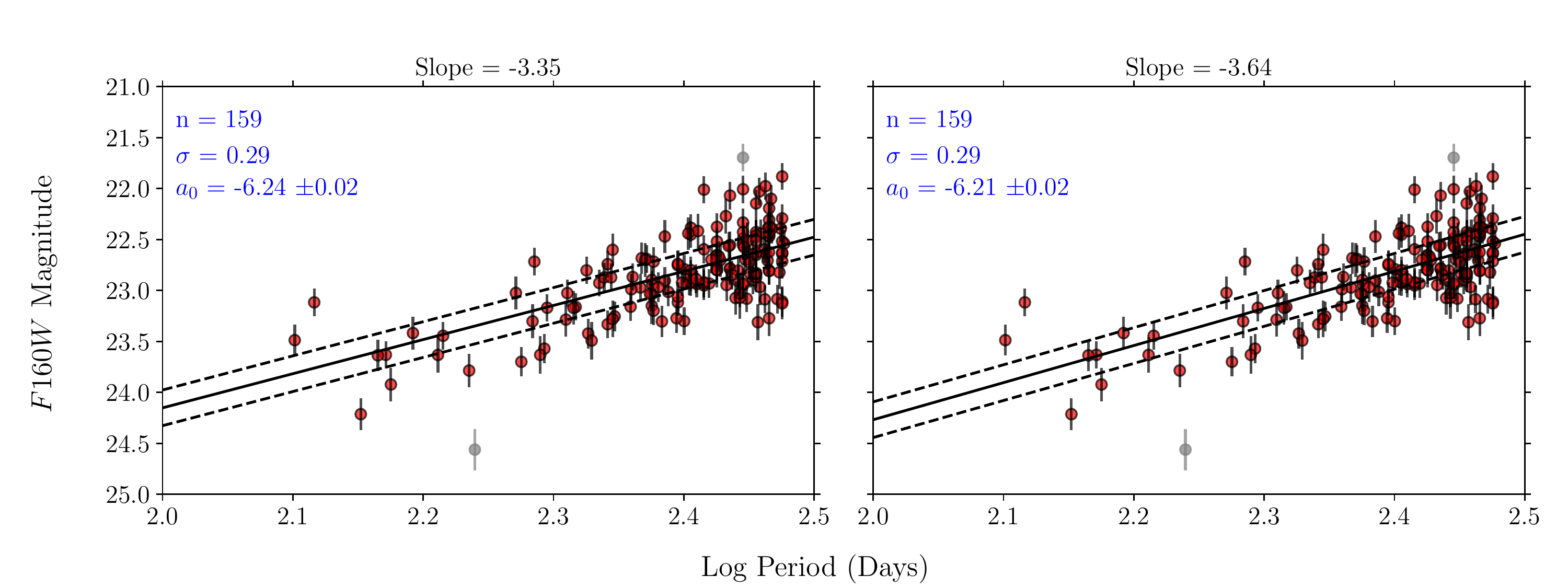}
  \caption{The linear PLR for NGC 4258 for the gold sample of Miras from H18 with the \emph{F160W} slope derived from the OGLE LMC O-rich Mira data (left) and the \emph{H}-band slope from \citet{Yuan17a}. H18 originally used only a quadratic PLR. Thus, we have refit the same data with a linear PLR to obtain a calibration for the linear Mira PLR. Red points were used in the fit and gray points were excluded from the clip via sigma-clipping. Dashed black lines show the $1\sigma$ dispersion. The error in the zeropoint only include the statistical errors.
   }
  \label{fig:plrlinear}
\end{figure*}

The uncertainty for each Mira mean magnitude (determined from the sine fit) is given by
\begin{align}
 \sigma_{\text{tot}} = \sqrt{\sigma_{\text{int}}^2 + \sigma_{\text{crowd}}^2},
\end{align}
where $\sigma_{\text{int}}$ is the intrinsic scatter of the Mira PLR and $\sigma_{\text{crowd}}$ is the measurement error given by the standard deviation of the recovered artificial star distribution.  We use an intrinsic scatter of $\sim 0.13$ mag derived from LMC observations of Miras. The DAOPHOT photometric errors are typically $ < 0.1$ mag, and are included in the artificial star results. The total uncertainty in most cases is dominated by the measurement error, as indicated by the artificial stars.




\begin{figure*}
\epsscale{1.2}
  \plotone{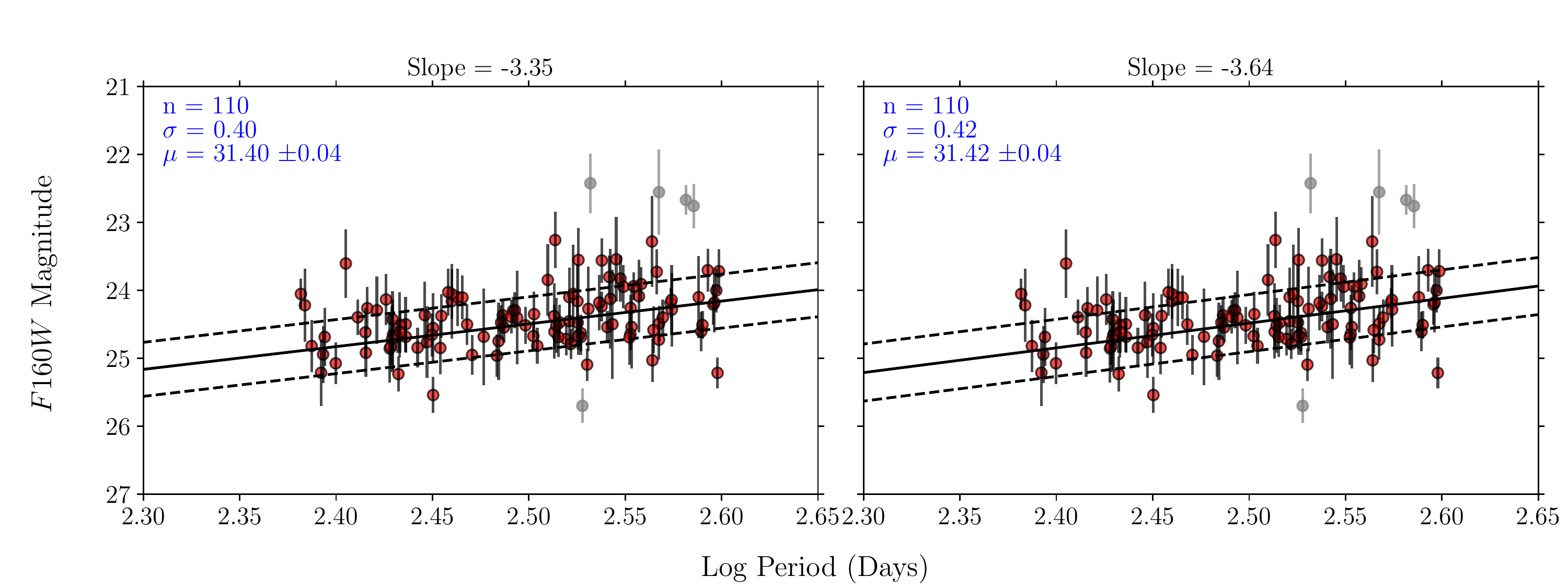}
  \caption{The final PLRs for NGC 1559 with the \emph{F160W} slope derived from the OGLE LMC O-rich Mira data (left) and with the \emph{H}-band slope from \citet{Yuan17a}. Red points were used in the fit and gray points were excluded via a $3\sigma$ clip, consistent with our artificial star tests. Dashed black lines show the $1\sigma$ dispersion. The uncertainties listed in the distance modulus include only the statistical errors.
   }
  \label{fig:plrlinear_double}
\end{figure*}

Using water megamasers orbiting the center of the galaxy, \citet{Reid19} obtained the most precise geometric distance to NGC 4258 to date, of $7.58 \pm 0.08$ (stat) $\pm 0.08$ (sys) Mpc. Thus, the distance modulus for NGC 4258 is $\mu = 29.398 \pm 0.0324$ mag. Using this we calculate an absolute calibration of $a_0 = \aone $ mag for the Mira linear PLR using the \emph{F160W} slope and $a_0 = \atwo $ mag for the $\emph{H}$-band slope (both containing systematic and statistical errors). These PLRs are shown in Figure \ref{fig:plrlinear}. 

We then use the absolute calibration from NGC 4258 to derive absolute distances to NGC 1559 of $\mu = \muone \pm \finalsyserr \text{ (sys)} \pm \finalstaterr \text{ (stat)}$ mag and $\mu = \mutwo \pm \finalsyserr \text{ (sys)} \pm \finalstaterr \text{ (stat)}$, respectively. The PLRs for NGC 1559 are shown in Figure \ref{fig:plrlinear_double}.

NGC 1559 was host to the normal SN Ia 2005df.  Its color-corrected, light-curve corrected peak magnitude in the Pantheon SN~Ia sample system \citep{Scolnic18} is $m_B = 12.14 \pm 0.11$ mag, resulting in a calibration of the fiducial SN~Ia absolute magnitude $M_B^0= -19.27 \pm 0.137 $ mag.  The uncertainty in $M_B^0$ is given by
\begin{align}
\label{eq:Msig}
\sigma_{M} = \sqrt{\sigma_{\mu 1559}^2 + \sigma_{\text{SNe}}^2},
\end{align}
where $\sigma_M$ is the uncertainty in $M_B^0$, $\sigma_{\mu 1559}$ is the uncertainty in the Mira-calibrated distance modulus to NGC 1559, and $\sigma_{\text{SNe}}$ ($=0.11$ mag) is the uncertainty in the corrected peak magnitude of SN 2005df. 
This agrees well with the fiducial value for SNe~Ia calibrated by Cepheids in 19 SN~Ia hosts (R16) of $M_B^0= -19.28 \pm 0.05$ mag based on the geometric calibrations of Cepheids in NGC 4258.

From the combination of the SN Ia absolute magnitude derived here and the intercept of the SN Ia Hubble diagram we can determine the value of the Hubble constant.

We derive $H_0$ using
\begin{align}
\log H_0 = (M_B^0 + 5a_B + 25)/5,
\end{align}
where $a_B$ is the intercept of the SN~Ia magnitude-redshift relation. 
The value for $a_B$ is adopted from R16, $a_B = 0.71273 \pm 0.00176$. 
To get the uncertainty in $H_0$, we sum in quadrature $\sigma_{M}$ and $5\sigma{a_B}$.

We find $H_0 = \HCfullp $ km\,s$^{-1}$\,Mpc$^{-1}$ with a measurement uncertainty of 6.3\% 
based on the masers in NGC 4258 as the sole geometric source of calibrating the luminosity of Miras.

We also calculate the distance modulus to NGC 1559 using only Miras with periods between 240 and 300 days. This is the period range of Miras in NGC 1559 that overlaps with the periods of Miras in NGC 4258. Using the subset of Miras from NGC 4258 in this period range to calibrate, we find that for the \emph{F160W} slope, we have $a_0 = -6.25 \pm 0.042$ mag. For the \emph{H}-band slope this becomes $a_0 = -6.21 \pm 0.042$ mag (statistical and systematic errors included in both). This gives us a distance modulus to NGC 1559 of $\mu = 31.36 \pm 0.10$ mag using the \emph{F160W} slope and $\mu = 31.36 \pm 0.10$ mag using the $H$-band slope, resulting in $H_0 = \HCpartp$ km\,s$^{-1}$\,Mpc$^{-1}$ for the former. Selecting a smaller period range increases our statistical error. 

As an alternative source of calibration of Mira luminosities we can use the LMC in lieu of NGC 4258.  The Miras from the OGLE-III sample of long-period variables in the LMC and our color terms yield an intercept (apparent) for the LMC sample of $\lmczpt \pm 0.01$ mag. Using the distance modulus to the LMC from detached eclipsing binaries of $\lmcdistmod$ mag \citep{Pietrzynski19}, we obtain an absolute calibration of $a_0 = \lmccal \pm 0.03$ mag in the \emph{F160W} bandpass. Using this calibration, we estimate a distance modulus to NGC 1559 of $\mu = 31.38 \pm 0.065$ mag. The calibration of SN 2005df using the LMC Mira sample results in $M_B^0= -19.24 \pm 0.128 $ mag and gives us $H_0 = \HCLMC$ km\,s$^{-1}$\,Mpc$^{-1}$.

\begin{figure}
\epsscale{1.2}
  \plotone{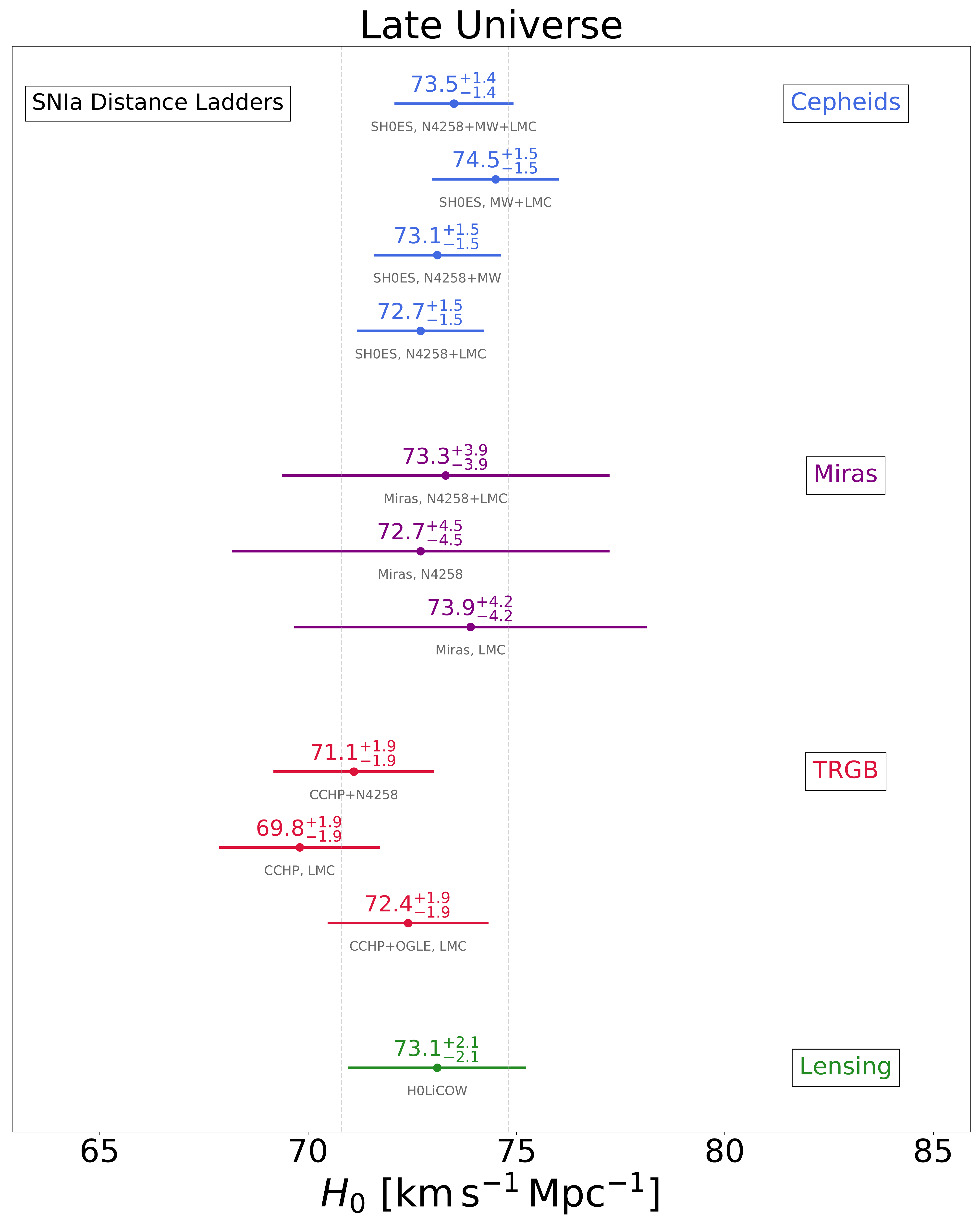}
  \caption{A summary of recently-published late-Universe $H_0$ measurements using different SN Ia distance ladders and anchors and their $1\sigma$ uncertainties. The set shown are those reviewed by \citet{Verde19}.  Featured are Cepheid results from R16 and R19 (in blue) with calibration from \citet{Reid19}, Mira results from this paper (in violet), TRGB from \citet{Freedman19} (CCHP), \citet{Yuan19} (CCHP+OGLE) and \citet{Reid19} (CCHP+N4258) (in red), and gravitational lensing time delays from \citet{Taubenberger19} used to calibrate SNe Ia (in green). For Cepheids, Miras, and TRGB, the combinations of anchors used are denoted in gray under the measurement points. Gray dashed vertical lines mark a range of 4 km\,s$^{-1}$\,Mpc$^{-1}$, $\pm 2$ km\,s$^{-1}$\,Mpc$^{-1}$ centered on the mean of the measurements. 
   }
  \label{fig:hclate}
\end{figure}

We combine the two anchors by taking a weighted mean of the distances to NGC 1559 using the LMC and using NGC 4258 as anchors. The uncertainty in the weighted mean distance is given by
\begin{equation}
\sigma_{\text{wm}} = \sqrt{\left( {\frac{1}{\sigma_{\text{LMC}}^2} + \frac{1}{\sigma_{4258}^2}}\right)^{-1}},
\end{equation}
where $\sigma_{\text{LMC}}$ is the uncertainty using the LMC as an anchor and $\sigma_{4258}$ is the uncertainty using NGC 4258 as an anchor. Combining the two anchors, $H_0 = \HCCombined $ km\,s$^{-1}$\,Mpc$^{-1}$ and $M_B^0= \absmagsne$ mag. 

All of these measurements of $H_0$ are within $1\sigma$ of the Cepheid-based values from R16 and R19, and they are dominated by the statistical uncertainty from the sole SN~Ia calibrator. Both the systematics inherent to the Mira PLR ($\sim 0.05$ mag) and the statistical error ($\sim 0.05$ mag) are subdominant to the random error ($\sim 0.11$ mag) of having only one calibrator. We are targeting three additional SN~Ia host galaxies that will eventually have both Cepheid and Mira calibrations of SNe. These additions to the local calibrator SN sample size will decrease the overall uncertainty on the Mira $H_0$ by a factor of 2 and decrease the random uncertainty from the the SN calibrators. 

Figure \ref{fig:hclate} shows a set of different approaches recently used to calibrate the SN Ia distance ladder and measure $H_0$ as recently reviewed by \citet{Verde19}.  As shown, the Mira calibration of SN Ia is in good agreement with the mean from Cepheid, TRGB and strong lensing.

The Mira sample in NGC 1559 contains the most distant Miras for which we have measured periods and magnitudes. They are also the first Miras to be used to calibrate the absolute luminosity of a SN~Ia, and thus the first Mira-based determination of $H_0$. With the absolute calibration from NGC 4258 on the \emph{HST} filter system, we can remove biases from converting from the ground-based filter magnitudes to \emph{HST} WFC3/IR magnitudes. 

In addition to showing that we can detect Miras in nearby SN hosts, we have shown that it is possible to find them in the halos of galaxies. This confirms that Miras can be used to increase the nearby SN sample by allowing us to look at edge-on galaxies in addition to early-type galaxies that are not actively star-forming.  

While newer space telescopes like \emph{JWST} may eventually surpass the ability of \emph{HST} to observe Miras in the NIR, \emph{HST} currently remains the best option for searching for Miras in SN host galaxies. \emph{JWST} offers advantages and challenges for observing Miras.  Its higher sensitivity and resolution are great advantages to reach more-distant hosts, but searching for time-variable phenomena poses challenges as \emph{JWST} has only two continuous viewing zones near the North and South ecliptic poles, where year-round observations would be possible. 

\section{Conclusions}
\label{sec:conclusions}

In H18, we calibrated the absolute magnitude of Miras using a sample observed with \emph{HST} WFC3-IR in the megamaser host galaxy NGC 4258. In this paper, we present a sample discovered in NGC 1559, use them to calibrate SN 2005df, and demonstrate that Miras can fulfill a role historically occupied by Cepheids as the second rung of the distance ladder. Now we are undertaking a program to calibrate the luminosity of SNe~Ia in other galaxies with Miras, which can yield an independent cross-check on the Cepheid-based $H_0$ measurements. 

To review, we identified $\sim 3000$ potential variable objects in 10 epochs of \emph{HST} NIR imaging of NGC 1559, and we used selection criteria designed to identify O-rich Miras to narrow this down to a final sample of 115 objects. We empirically corrected for sample contamination by C-rich Miras using the LMC as a model. 

Potential systematics that affect our measurement include the reddening, color, bias due to crowding, and uncertainty in the slope of the PLR. Our estimate of the zeropoint is dominated by the systematic error, since we are only fitting a single parameter to the PLR. Reddening should be negligible based on small foreground extinction and location of the Miras in the halo of NGC 1559, away from the most crowded regions. However, for the distance scale, we are concerned with the differential extinction between NGC 4258 and NGC 1559, which we estimate conservatively to be $\sim 0.04$ mag.

Finally, we determine a distance to SN 2005df host NGC 1559, the first estimate of distance to a such a host with Miras. We use the absolute calibration from NGC 4258 to anchor the Mira distance scale and measure a value of $H_0$, $\HCfullp$ km\,s$^{-1}$\,Mpc$^{-1}$, in good agreement with the Cepheid-calibrated $H_0$ to within $1\sigma$ of the uncertainties. In addition, the values of $H_0$ for the subset of the data at the same period range as the Miras in NGC 4258, $\HCpartp$ km\,s$^{-1}$\,Mpc$^{-1}$, and (adding the LMC as an additional anchor) our preferred value of $\HCCombined$ km\,s$^{-1}$\,Mpc$^{-1}$, are also consistent with the Cepheid-calibrated $H_0$. 

This paper demonstrates that Miras can be used as calibrators for local SNe~Ia. The ubiquity of Miras means that they can be found in a wider range of local SN~Ia hosts, thereby increasing the sample of possible calibrators, and also potentially making the results more representative of the Hubble-flow sample. 

\section*{Acknowledgements}
\label{sec:ack}

We are grateful to John Soltis for providing Figure \ref{fig:hclate}, the compilation of recent $H_0$ measurements. We would also like to thank the anonymous referee for their helpful comments. 

Support for this work was provided by the National Aeronautics and
Space Administration (NASA) through {\it HST} program GO-15145 from
the Space Telescope Science Institute (STScI), which is operated by
AURA, Inc., under NASA contract NAS 5-26555. A.V.F. is also grateful
for financial assistance from the TABASGO Foundation, the Christopher
R. Redlich Fund, and the Miller Institute for Basic Research in
Science (UC Berkeley).  P.A.W. thanks the South African NRF for a
research grant.

\bibliographystyle{apj}
\bibliography{ms_agr}

\end{document}